\documentclass[twocolumn]{aastex631} 
\usepackage{amsmath, amsthm}
\usepackage{CJK}
\allowdisplaybreaks[1]

\submitjournal{Astrophysical Journal Letters}

\shorttitle{The light curve of wind-reprocessed TDEs}
\shortauthors{Mockler et al.}
\graphicspath{{./}{figures/}}

\begin{document}
\begin{CJK*}{UTF8}{gbsn}
\title{The Light Curve of Wind-Reprocessed Tidal Disruption Events}


\correspondingauthor{Brenna Mockler}
\email{bmockler@carnegiescience.edu}

\author[0000-0001-6350-8168]{Brenna Mockler}
\affiliation{The Observatories of the Carnegie Institution for Science, Pasadena, CA 91101, USA}
\affiliation{Department of Physics and Astronomy, University of California, Davis, CA 95616, USA}

\author[0000-0003-4307-0589]{David Khatami}
\affiliation{Lawrence Livermore National Laboratory, Livermore, CA 94550, USA}

\author[0000-0002-5981-1022]{Daniel Kasen}
\affiliation{Department of Astronomy and Theoretical Astrophysics Center, University of California, Berkeley, CA 94720, USA}
\affiliation{Nuclear Science Division, Lawrence Berkeley National Laboratory, Berkeley, CA, 94720, USA}

\author[0000-0003-2868-489X]{Xiaoshan Huang}
\affiliation{California Institute of Technology, TAPIR, Pasadena, CA 91125, USA}

\author[0000-0001-6806-0673]{Anthony L. Piro}
\affiliation{The Observatories of the Carnegie Institution for Science, Pasadena, CA 91101, USA}

\begin{abstract}

The source of the optical/UV emission in tidal disruption events (TDEs) remains an enduring question in the field. Connecting the observed emission to the source is critical for both our understanding of these transients and for using TDEs to study the efficiency of super-Eddington accretion and black hole growth. To explore this connection, we ran time-dependent 1D radiation hydrodynamic simulations of TDE emission with the {\tt Sedona} monte carlo radiative transfer code, focusing on the reprocessing paradigm. Our simulations follow a compact, evolving X-ray and EUV bright source and surrounding reprocessing outflow over multiple months, using luminosities and mass flow rates consistent with hydro simulations of tidal disruptions. We determine the efficiency of reprocessing as a function of time in this dramatically changing environment and reproduce key observables including timescales, luminosities, and color evolution. Notably, we see a strong wavelength-dependence in the emission timescale due to reprocessing effects. Early on there is an X-ray flare which quickly fades as material builds up and obscures the hot source. At the same time, the optical/UV luminosity begins to rise. Though the optical/UV light curve has a similar shape to the bolometric light curve, the optical peak is offset by $\sim$3 weeks from the bolometric peak due to the time required to build up the reprocessing layer. This implies that early time, high energy emission may be missed for TDEs discovered in optical surveys, and the initial disruption and mass return time to the black hole may occur earlier than optical light curves suggest.

\end{abstract}


\section{Introduction} \label{sec:intro}

Tidal disruption events provide us with an avenue to study the supermassive black holes (SMBHs) and stars that live at the very centers of galaxies, as well as SMBH accretion that varies on human timescales., 
During these transients, the supermassive black holes that disrupt the stars are often fed near or above their Eddington limits \citep[][]{Rees:1988a, van_velzen_optical-ultraviolet_2020, mockler_energy_2021, hammerstein_final_2023-1}, probing the limits of black hole growth.  Recently, JWST has uncovered evidence of potential highly accreting supermassive black holes in the early universe that have stretched our understanding of black hole formation and growth 
\citep[e.g.][]{matthee_little_2024}. Most local active galactic nuclei (AGN) are accreting far below their Eddington limits \citep[e.g.,][]{jones_intrinsic_2016, wang_identifying_2024, blanton_eddington_2025}, and so tidal disruption events are critical for studying SMBH growth limits in detail at lower redshift.

The process of forming a disk from the stellar debris and feeding a black hole near the Eddington limit is messy, and past work has predicted significant material will be ejected, first from the shocks that circularize material, and later by radiation pressure from the highly accreting disk \citep[][]{jiang_prompt_2016, bonnerot_first_2021, steinberg_origins_2022, huang_bright_2023, Dai:2018}. All of this material can obscure the forming disk \citep[][]{steinberg_origins_2022, price_eddington_2024, hu_optical_2024, huang_x-ray_2025}, and means that radiation transport simulations are necessary to connect the emission from regions close to the black hole to the luminosity that escapes at much larger radii \citep[e.g.,][]{roth_X-ray_2016}.

One of the open questions in TDEs is the origin of the optical and UV emission that is ubiquitous in the optically discovered TDEs that dominate the observed TDE population \citep[e.g.,][]{van_velzen_optical-ultraviolet_2020, hammerstein_final_2023-1, yao_tidal_2023-1}. The reprocessing of high energy emission originating from close to the black hole by surrounding gas and outflows has often been put forward as a means to explain the lower energy emission together with the varying levels of X-ray flux observed during these events \citep[e.g.,][]{Strubbe:2009a, roth_X-ray_2016, Dai:2018}.

While significant progress has been made on the hydrodynamic simulations that track the disruption of the star \citep[see for example,][]{guillochon_hydrodynamical_2013, law-smith_stellar_2020, ryu_tidal_2020} and the initial shocks that dissipate energy and redistribute angular momentum to circularize material into a disk \citep[][]{jiang_prompt_2016, bonnerot_first_2021, huang_bright_2023}\footnote{Or do both! For example, \citealt{steinberg_origins_2022, ryu_shocks_2023, price_eddington_2024-1, martire_wind-mediated_2025}. }, there have been much fewer attempts focused on modeling the frequency, and time dependent radiation transport of emission from the inner regions to compute detailed light curve and spectral evolution and compare with observations. \citet[][]{huang_x-ray_2025} produced one of the first multi-group 3D radiation hydrodynamics simulations of the early phases of a TDE, however given the computational expense involved, they were only able to run the multi-group run for 10-30 minutes of physical time. Even more recently, \citet[][]{giron_multigroup_2026} produced a multi-group 3D radiation hydrodynamics simulation of a TDE around an intermediate mass black hole that ran for $\sim 5$ days. Both of these works are revolutionizing our understanding of the optical emission in TDEs, however they still only cover a small fraction of the TDE evolution and don't model the emergent spectrum. The majority of the simulations that do model the emergent spectra are essentially time-independent -- they either post-process hydrodynamical simulations \citep[][]{ thomsen_dynamical_2022-1}, or assume steady-state hydro conditions for their runs \citep[][]{roth_X-ray_2016, parkinson_optical_2022}. 
However, in a tidal disruption event, the gas flow to the black hole changes dramatically with time \citep[][]{guillochon_hydrodynamical_2013}, and gas conditions can change dramatically over the timescale it takes photons to advect and diffuse out of the gas. During the TDE, 
radiation pressure can drive outflows \citep[][]{Dai:2018}, and adiabatic losses from gas expansion will change the photon energetics \citep[][]{roth_what_2018} as well.
This is similar to the behavior described more generally for `wind-reprocessed' transients in \cite[][]{piro_wind-reprocessed_2020} and \cite[][]{calderon_moving-mesh_2021}. 
Therefore, modeling the dynamical coupling between the radiation and gas is critical to paint an accurate picture of how the observed light curves of these transients connect to the underlying physics. 

In what follows, we explore whether the reprocessing of X-ray-bright emission from near the black hole by surrounding gas can explain both the bright optical and UV luminosities and the varying levels of X-ray emission seen in TDEs. We first explain the motivation for modeling TDE emission through a wind-reprocessing framework in Section~\ref{sec:reprocessing_background}. Then, in Section~\ref{sec:methods}, we describe how we use the {\tt Sedona} radiation transport code \citep[expanded to include 1D moving-mesh hydrodynamics,][]{kasen_time-dependent_2006, khatami_landscape_2023}, to model the rise, peak, and initial decline of a TDE.
We show in Section~\ref{sec:observables} that the light curve and spectral evolution obtained from our simulations are consistent with observed events, but notably find that the peak of the bolometric light curve pre-dates the optical peak by weeks due to the time it takes to build up an optical thick reprocessing layer. This implies that by the time many events are discovered in optical TDE surveys, their \textit{bolometric} luminosity may actually be declining (even as their optical/UV emission rises). Relatedly, we predict that some events may produce an X-ray flare well before optical peak, similar to what was observed in \citet[][]{malyali_transient_2024}. We also connect the hydrodynamic properties and ionization state of the gas reprocessing the emission to the actual reprocessed spectra and light curves. We find that the opacities and optical depths changes dramatically over the course of the simulation, and that it is the outer layers of the wind that determine the opacity to X-rays, as this is where the ionization states of metals are lowest. This means that even when X-rays are not observable, they can irradiate a significant fraction of the envelope, only getting absorbed in the outer regions of the wind.
Finally, in Section~\ref{sec:discussion} and ~\ref{sec:summary} we discuss the main takeaways from our work, as well as caveats, motivation for future observations, and next steps.

\section{Reprocessing in TDEs}\label{sec:reprocessing_background} 
\begin{figure}[ht!]
\begin{center}
\includegraphics[width = \columnwidth]{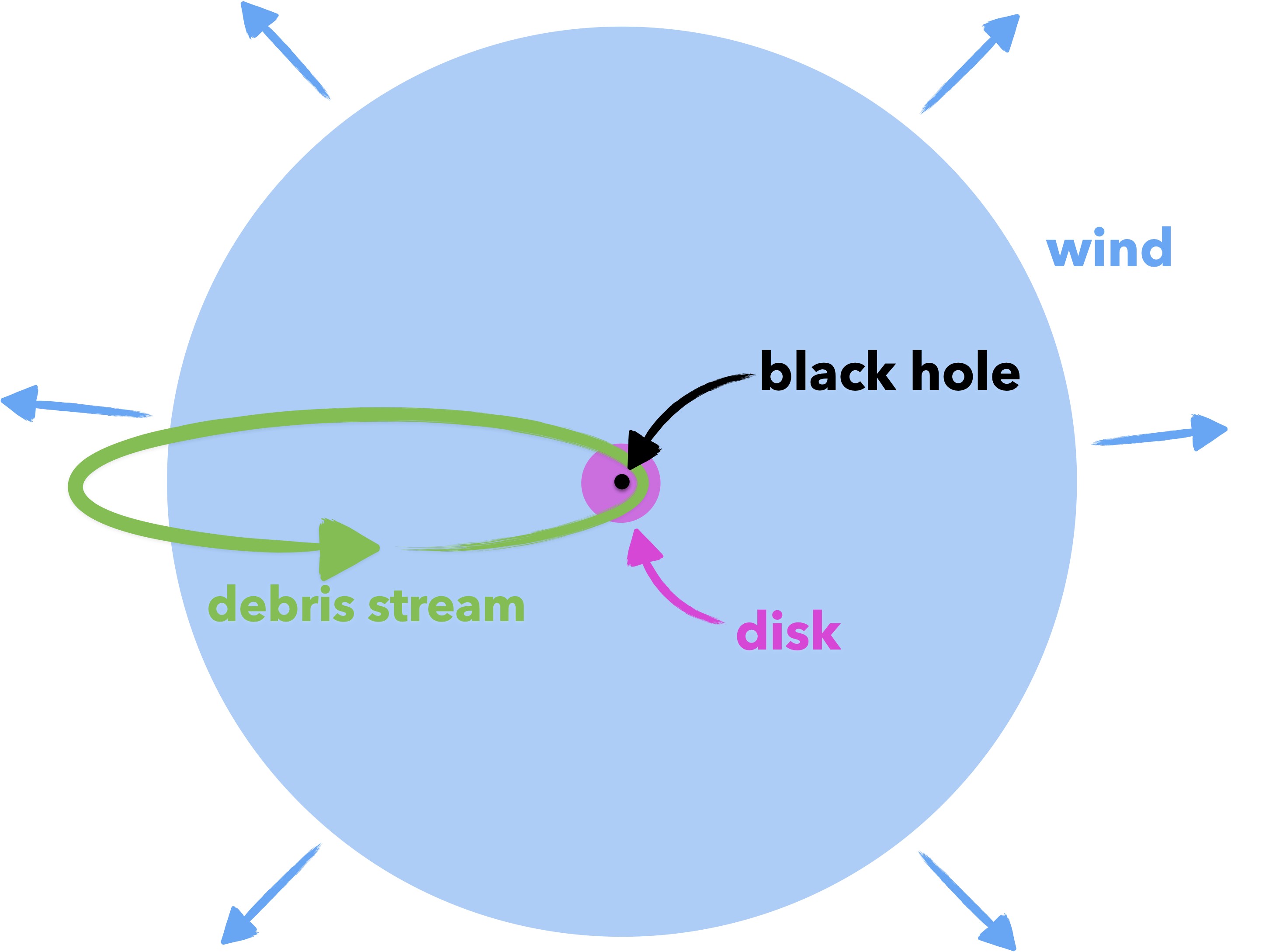} 
\end{center}
\caption{This cartoon depicts the main components of a wind-reprocessed TDE. In our simulation, we are assuming a hot X-ray source at the size scale of the forming disk, and reprocessing its emission through winds outflowing from the same radius. As this is a 1D simulation, we do not include the stream, however we use the orbital properties of the disruption to determine the circularization radius and mass and luminosity evolution.
\label{fig:TDEcartoon}
}
\end{figure}

This work is motivated by the multiple lines of evidence we see for reprocessing in TDEs. Tidal disruption events were originally predicted to be X-ray bright, but most are now discovered by optical surveys \citep[e.g.,][]{van-Velzen:2011a, Holoien:2014a, hammerstein_final_2023-1}. When X-rays are observed, the optical to X-ray ratio is generally greater than one \citep[and between 10-1000 near peak optical for most events][]{guolo_systematic_2023}. Unlike AGN, these X-rays are usually very soft, emitting most of their energy $< 1 {\rm keV}$, with hardness ratios $<0$ \citep[][]{Auchettl:2017a, Auchettl:2017b, guolo_systematic_2023}.
The average blackbody radii estimates for the optical/UV emission are a few $\times 10^{14}$ cm -- more than $1000\times R_g$ for a $10^6 M_\odot$ black hole \citep[e.g.,][]{hung_revisiting_2017, hammerstein_final_2023-1}. Additionally, correlations between the spectral line ratios of, for example, He and H and the size of the fitted blackbody radius of the photosphere also indicate that the optical emission is being reprocessed at various radii within the photosphere \citep[][]{van_velzen_seventeen_2021, charalampopoulos_detailed_2022, hinkle_examining_2020}. This is consistent with TDE simulations -- both of the initial circularization process \citep[][]{jiang_prompt_2016, bonnerot_first_2021, huang_bright_2023, steinberg_origins_2022, price_eddington_2024}, and then eventually of the puffy disk  \citep[][]{dai_unified_2018,thomsen_dynamical_2022-1, andalman_tidal_2022}-- which find significant material surrounds the black hole and at large radii, enshrouding emission both from the initial stream collisions and the forming disk. Finally, over the past few years, more and more TDEs have been discovered to have radio emission consistent with outflows moving at $\sim 0.01-0.1c$ \citep[][]{cendes_ubiquitous_2023, goodwin_at2019azh_2022, goodwin_radio-emitting_2023,  alexander_multi-wavelength_2025}.

In this work we study how X-ray bright source emission originating on size scales comparable to the circularization radius is reprocessed by material that is initially ejected\footnote{though not necessarily unbound}, from the black hole. We have included a simple schematic of the scenario we are testing in Figure~\ref{fig:TDEcartoon}, though we note that in this (1D) work we are not simulating the debris stream or the hydrodynamics of the disk. This allows us to remain agnostic as to whether the source of the X-ray emission is primarily shocks from the disk formation process or accretion through the disk itself, though we do assume that it originates from the size scale of the forming accretion disk and therefore is an example of the ``prompt'' circularization scenario. This scenario is expected to occur when the initial stream collisions happen close to the black hole at high velocities \citep[e.g. see ][]{huang_bright_2023}, or if the nozzle shock is effective at liberating energy (see \citealt[]{steinberg_origins_2022} but also \citealt[]{hu_converged_2026}). Observationally, detections of early time X-ray emission \citep[][]{malyali_transient_2024, guolo_systematic_2023}, double-peaked line profiles \citep[][]{short_tidal_2020, hung_double-peaked_2020, wevers_elliptical_2022}, and line emission requiring high ionizing fluxes \citep[e.g. events with Bowen lines][]{blagorodnova_iptf16fnl_2017, Leloudas:2019} point to this ``prompt'' circularization scenario being relevant for at least a subset of tidal disruption events.

\section{Methods}\label{sec:methods}

Our radiation hydrodynamics simulations are run with the radiation transport code {\tt Sedona}. We use the implicit Monte Carlo method from \citet[][]{roth_monte_2015} and updated in \citet{khatami_physics_2024} with the recently-implemented moving-mesh hydrodynamics from \citet{khatami_landscape_2024}. Our implementation solves the radiation transport equations described in Section 4.2 of \citet{khatami_physics_2024} (Equations 4.1 - 4.36). We also assume an ideal gas equation of state for the hydrodynamics with $\gamma = 5/3$, but note that the effective equation of state is dependent on whether or not the gas is radiation dominated in a particular cell and therefore varies between $\gamma = 5/3$ and $\gamma = 4/3$. We use a reflecting boundary condition at the inner boundary, appropriate for a hot, dense source.
The simulations are run in 1D spherical symmetry due to the computational expense of solving for the time-dependent radiation transport + hydrodynamics. 

Because the focus of these simulations is to model the time-dependent SED and not the shapes and magnitudes of individual lines, the radiation transport calculations are solved assuming that the level populations and ionization states in the gas are in local thermodynamic equilibrium (LTE). Therefore, the level populations obey the Boltzmann distribution and the ionization states are described by the Saha equation \footnote{We include a comparison with steady-state non-LTE calculations in Appendix~\ref{sec:nLTE_comparison}.}. However, we do not assume that the gas and radiation are in thermal equilibrium, and instead solve for the energy of the gas at each timestep. We use on-the-fly LTE level populations and opacities calculated in {\tt Sedona} using screened hydrogenic atomic data \citep[][]{chung_generalized_2016}. We also use a simplified treatment of line absorption, using a line absorption efficiency parameter `$\epsilon$' for the bound-bound transitions, similar to the approach in the two-level atom approximation \citep[][see Section~\ref{sec:opacities} for more details]{mihalas_stellar_1978}. This `$\epsilon$' is effectively a global ratio of the absorption coefficient for bound-bound transitions to the coefficient for electron scattering ($\epsilon = \alpha_{\rm bb}/\alpha_{\rm es}$). While the true value of `$\epsilon$' is  unique to each line, here we approximate it with a single value for each simulation, following the approach of \citet{kasen_time-dependent_2006}. It enters into the source function for the lines as described by Equation 7 in \citet{kasen_time-dependent_2006} \citep[see also][]{nugent_synthetic_1997}.

Our hydro simulation grid has 200 cells that are initiated with log-spacing, an inner boundary of $10^{13}$ cm (the size of the circularization radius for our system, or approximately $20 \times R_g$), and an initial innermost (smallest) zone size of $10^{10}$ cm $= 0.14  R_\odot$. The outer boundary of the simulation is initially at $1.26 \times 10^{13}$ cm and eventually expands to $3.94\times 10^{15}$ cm over the course of the simulation. We note that the size of the cells expand in time with the wind due to the moving mesh. We use a log-spaced frequency grid covering X-ray to mid-infrared wavelengths ($3-3 \times 10^5 \AA$), which results in a frequency grid size of 1158.

\subsection{Simulation setup}\label{sec:simsetup}

We model the reprocessing of emission for a fiducial TDE with the following parameters: 

\begin{itemize}
    \item $M_h = 3 \times 10^6 M_\odot$
    \item $M_* = 1 M_\odot$
    \item $\beta = R_t/R_p = \beta_{\rm crit} = 1.85$,
\end{itemize}

Where $\beta_{\rm crit}$ is the minimum impact parameter for full disruption for the star\footnote{The star is modeled as a polytrope with polytropic index $\gamma = 4/3$.}. Our mass fallback rate is from hydrodynamic simulations from \cite{guillochon_hydrodynamical_2013}, and is also available through the {\tt MOSFiT} TDE model \citep[][]{Guillochon:2018, mockler_weighing_2019}. The peak mass fallback rate is $1.23 \times 10^{26}$ g/s or $1.95 M_\odot$/year. For this initial simulation we use a constant composition and include hydrogen (mass fraction = 0.7), helium (mass fraction=0.29), and oxygen as a representative metal (mass fraction = 0.01). We choose oxygen as the representative metal because it is generally the dominant absorber at soft X-ray wavelengths, and note that including other metals may increase the opacity at high energies (e.g. in the X-ray and far UV), but would be unlikely to have a large effect on the UV/optical continuum while the luminosity is high and the wind is highly ionized \citep[see][]{roth_X-ray_2016}.
We do note that in an actual TDE, the metallicity is likely to be time-dependent, with material from the outer layers arriving at early times and material from the core arriving after peak \citep{Gallegos-Garcia:2018, law-smith_stellar_2020}. This means the material in the outflow at early times is likely to be at lower metallicities compared to the material in the outflow after the peak of the fallback rate \citep[unless the star is post-main sequence,][]{mockler_tidal_2024}. We plan to explore the effect of this particular on line emission in future work, however, given that only a small metal fraction is required to greatly increase the opacity to X-rays, we doubt this would dramatically change the overall SED evolution in the work presented here.
 
We assume the wind is launched from the radius of the disk at velocity $v_{\rm wind} = 0.01c$, and use $10^{13}$ cm as the effective size of the disk (this is approximately the circularization radius for our simulation parameters $ = 2 \times R_p = 7 \times 10^{12}$ cm). The outflow velocity is chosen based on observations of TDE outflows measured in the radio \citep[][]{cendes_ubiquitous_2023, alexander_multi-wavelength_2025} and blueshifts measured in spectra \citep[][]{nicholl_outflow_2020, charalampopoulos_detailed_2022}. It is also comparable to outflow velocities found in 3D hydro simulations of TDEs \citep[][]{bonnerot_first_2021, thomsen_dynamical_2022-1, huang_x-ray_2025, steinberg_stream-disk_2024}\footnote{It is comparable at intermediate viewing angles, this is slower than outflows near the poles/jets.}.
We assume that $75\%$ of the mass that returns to the black hole is launched in an outflow\footnote{We note that this mass is not unbound at these velocities.}, again informed by simulations such as \citet[][]{bonnerot_first_2021, price_eddington_2024, steinberg_stream-disk_2024, huang_bright_2023, huang_pre-peak_2024, huang_x-ray_2025}, which find $\sim 50\%$ or more of the fallback material is initially launched into non-relativistic outflows through either shocks or accretion. In fact, our time-dependent density profiles (see Figure~\ref{fig:hydro}) cover nearly the same range in parameter space as the spherically averaged density profiles in \citet[][]{huang_x-ray_2025} when comparing relevant time ranges.  \citet[][]{huang_x-ray_2025} is a 3D radhydro simulation with similar parameters but run for less time and without as detailed a treatment of radiation transport.

We start the simulation at $10\%$ of the peak fallback rate (1 week after the time of first fallback), and distribute the material that has previously returned around the black hole using the same mass fraction in the wind ($75\%$) and assuming the same launch velocity ($0.01c$) as we use throughout the simulation. This helps us initiate the simulation stably, and amounts to only $0.75 \times 0.0027 M_\odot$ of pre-distributed material. It also is likely a more accurate representation of the scenario seen in 3D simulations, where even in the prompt circularization scenario (for example, from strong stream collision shocks), it can take a week or longer for circularizing gas to form a hot source near the black hole \citep[][]{andalman_tidal_2022, huang_x-ray_2025}. 

To determine our source luminosity, we then assume an overall $1\%$ efficiency of mass to emitted luminosity from the mass fallback rate. This corresponds to a $4\%$ mass to energy efficiency of the mass that is presumed to actually make it to the black hole ($25\%$ of the total mass fallback rate) and a source luminosity that is $\sim 2.5 \times$ the Eddington limit at peak (the observed luminosity is lower, see Section~\ref{sec:energetics}).  

\begin{equation}
    L_{\rm source} = \epsilon_{\rm rad}f_{\rm fb}  \dot{M}_{\rm fb} c^2
\end{equation}
    
Where $\epsilon_{\rm rad} = 0.04$ and $f_{\rm fb} = 0.25$. This value of the mass-to-energy efficiency ($\epsilon_{\rm rad}f_{\rm fb} = 0.01$) is in the range of predicted total mass-to-energy efficiencies for near-Eddington accretion \citep[e.g][]{Dai:2018,jiang_super-eddington_2019, zhang_radiation_2025}, or from shocks near $R_t$ \citep[][]{jiang_prompt_2016, huang_pre-peak_2024}. Our source luminosity SED is then a blackbody at constant radius ($R_{\rm source} = 10^{13}$ cm) that evolves with time as determined by the evolving luminosity. While we recognize that assuming the source luminosity follows the fallback rate is a very rough approximation \citep[and certainly not accurate during the late-time plateau phase observed in many events, see][]{van_velzen_late-time_2019}, we use it as a first step given that our focus here is on the reprocessing of said emission. We plan in the future to take source luminosities from 3D hydro simulations, however there is still significant variance in the literature on what these source luminosities would look like even for simulations with similar parameters. 

The evolving source temperature is determined by the blackbody given by the source radius and the evolving luminosity as described above. 
However, back-heating of the gas near $r_{\rm in}$ due to the high optical depth of the wind causes the temperature at $r_{\rm in}$ to quickly re-calibrate to a higher value \citep[e.g. Equations 4 - 7 of ][]{Roth:2016a}, such that: 

\begin{equation}
    T(r_{\rm in}) \sim \Big (\frac{\tau_{\rm es}}{4} \frac{L_{\rm source}}{4 \pi \sigma_{\rm SB} r_{\rm in}^2}\Big)^{1/4}
\end{equation}

Where $\tau_{\rm es}$ is the optical depth to electron scattering through the gas around the source. 

We find that the peak temperature of the input SED does not have a strong effect on the output optical/UV SED, which is determined instead by the efficiency of reprocessing (though it will of course affect the observed X-ray emission). 

Finally, we assume a fiducial line absorption efficiency parameter of $\epsilon = 0.1$ (this provides a conservative estimate of the optical/UV luminosity, as higher values of $\epsilon$ redistribute more energy from shorter wavelengths to longer wavelengths) however we show in Appendix~\ref{sec:lineabsorption} that varying this parameter by factors of a few in either direction does not change the optical/UV continuum significantly (see also Section~\ref{sec:opacities}).

\section{Observables}\label{sec:observables}

Here we describe the results of our monte carlo radiation transport + 1D hydrodynamics simulations modeling the rise, peak, and decline of an optical/UV bright TDE. First, we analyze the resulting light curves and spectra and compare with properties of observed events. We then discuss how their evolution depends on the time-dependent properties of the source and surrounding outflow, particularly the ionization state and opacity of the gas. Finally, we summarize the energetics of the simulation and compare with analytical calculations.

\begin{figure*}[ht!]
\begin{center}
\includegraphics[width = 0.75\textwidth]{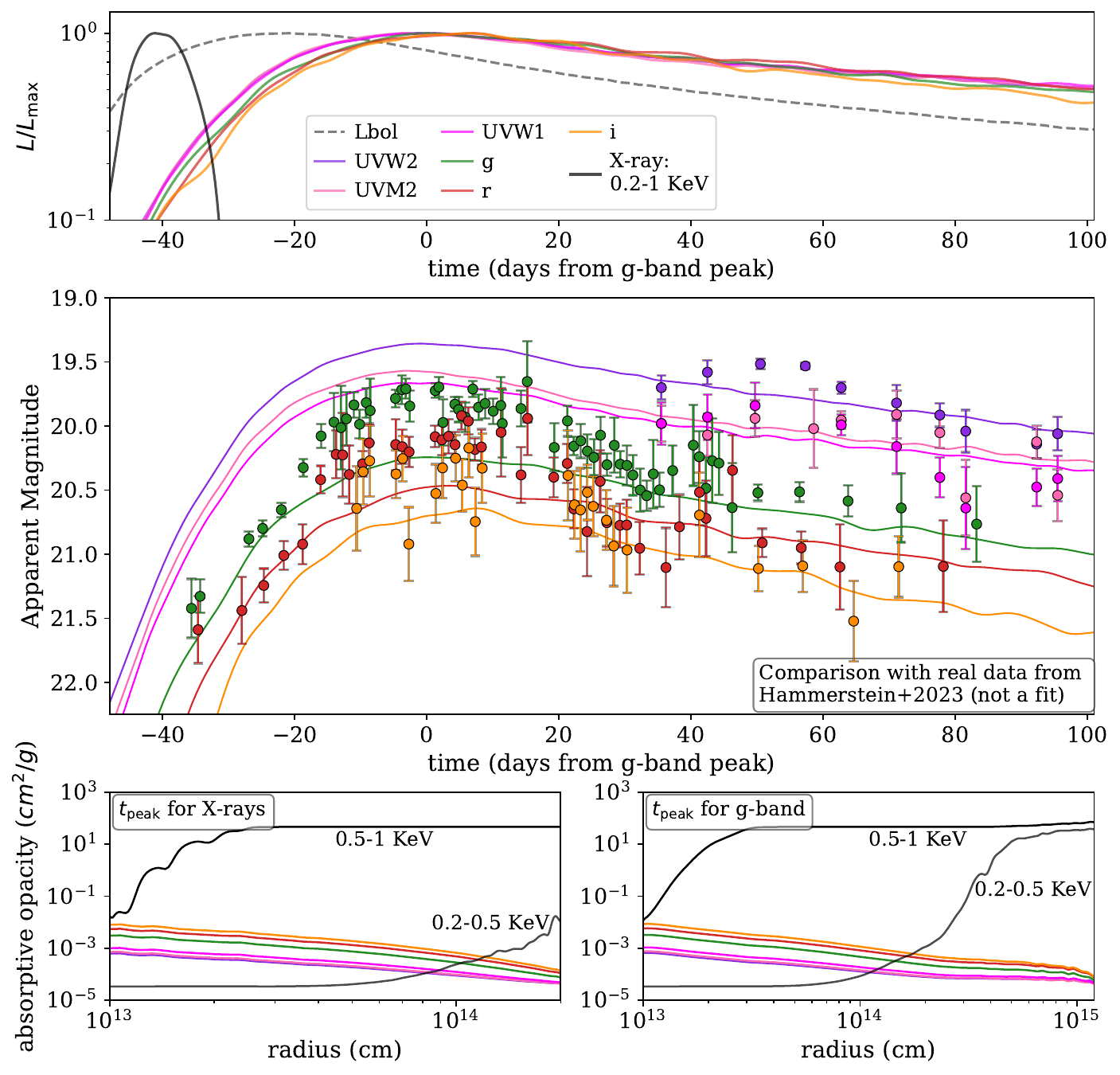} 
\end{center}
\vspace{-0.5cm}
\caption{We simulate the rise, peak, and initial decay of a TDE using MCRT + 1D hydrodynamics. {\bf Top Plot:} We plot the intrinsic optical/UV evolution, scaled to the peak luminosity in each band$^a$. We find that higher energy bands peak earlier, and UV/optical emission peaks weeks after the bolometric luminosity. X-rays peak before the bolometric luminosity, then get absorbed by the growing outflow which in turn produces the UV and optical emission that peaks later. Note that this is a 1D calculation, and therefore including viewing angles with lower optical depths would likely lead to different behavior of the X-ray emission in particular. {\bf Middle Plot:} We compare our optical through UV evolution to a real TDE (AT2020ocn) with a similar black hole mass ($4 \times 10^6 M_\odot$ from the $M-\sigma$ relation) and UV/optical blackbody parameters. We plot the apparent magnitude of our simulations at z=0.07 (time is in the observer frame) and assume a Milky Way hydrogen column density $n_{h, mw} = 10^{20} \rm cm^{-2}$ \citep[from][]{schlafly_measuring_2011} to compare with the light curve of AT2020ocn \citep[from][]{hammerstein_final_2023-1}. 
{\bf Bottom Plots:} We plot the weighted average of the absorptive opacity over the different bands (given the band transmission curves used in the top and middle plots) as a function of radius in the simulation. We split X-rays into harder and softer energy ranges. The {\it left} plot is at X-ray peak ($= 7$ days into the simulation), the {\it right} plot is at g-band peak (= $49$ days). The opacity to X-rays between $0.2-0.5$ keV increases dramatically at the outer edge of the simulation where gas is cooler and oxygen is in lower ionization states. By g-band peak, the opacity to all X-ray photons is orders of magnitude higher than the opacity to UV/optical photons.
\label{fig:lcs}
}
$^a${\footnotesize We use UV bands from Swift and optical bands from Palomar/ZTF. Our X-ray bands have a flat throughput$=1$ over the relevant energy ranges.}
\end{figure*}

\subsection{Luminosity evolution}

\begin{figure}[ht!]
\begin{center}
\includegraphics[width = 0.9\columnwidth]{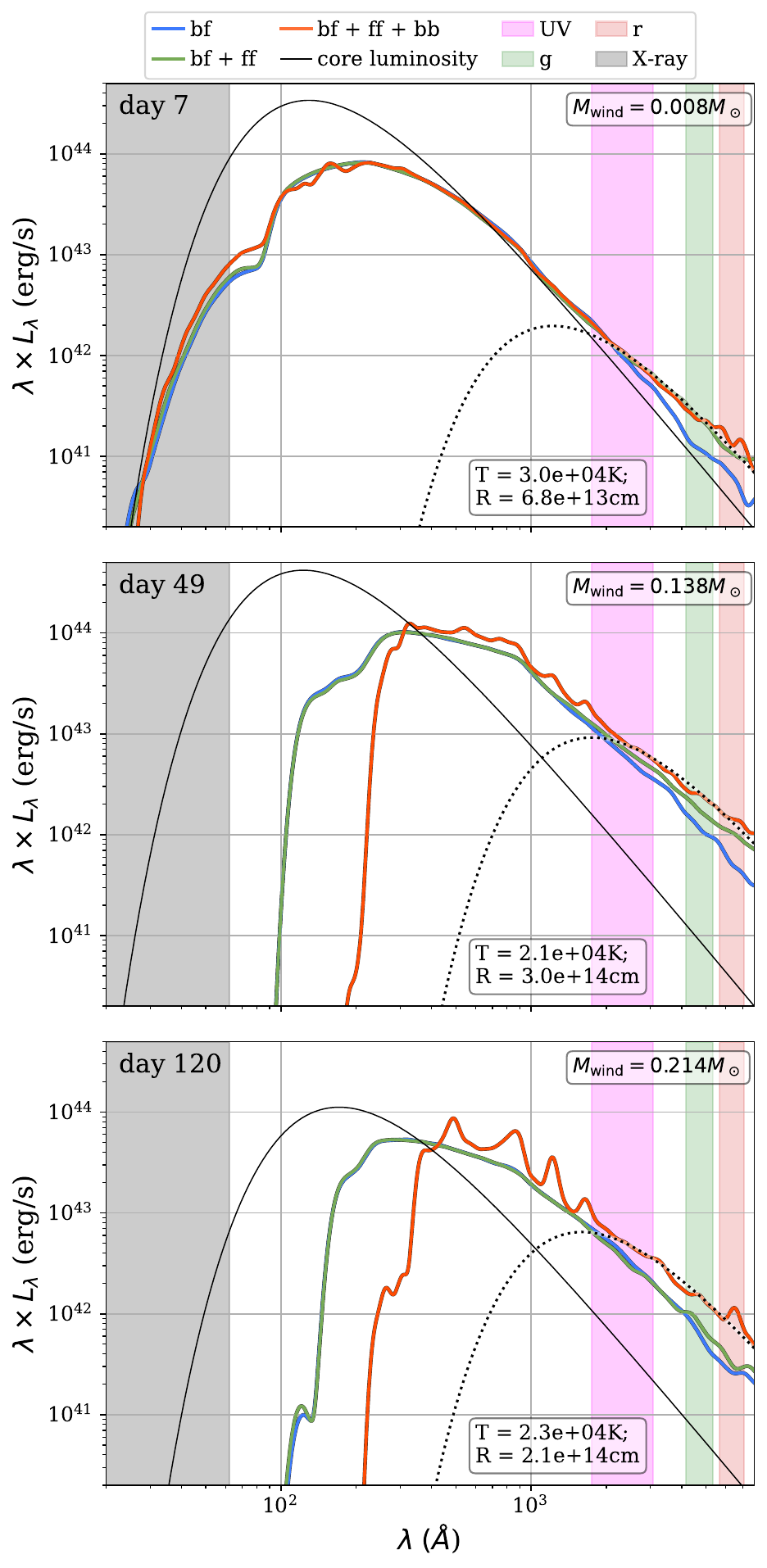} 
\end{center}
\caption{ The blue, green, and orange lines show the output SED using 3 different opacity prescriptions -- bound-free (blue), bound-free + free-free (green), and bound-free + free-free + bound-bound (orange). All three also include electron scattering opacities. The top plot shows emission early on the rise at 7 days into the simulation, the middle plot shows emission at 49 days (near optical/UV peak), and the bottom plot shows emission at 120 days, on the decline. The thin black solid lines show the source luminosity at these times, and the dotted black lines show example blackbody SEDs that match the emission relatively well in observable UV and optical bands, emphasizing the likelihood of underestimating the true luminosity when fitting blackbodies to observations.  The shaded black, magenta, green, and red vertical regions represented the approximated band range of eRosita, Swift UV, and Palomar g \& r bands. 
\label{fig:SEDevolution}
}
\end{figure}

We find that, in general, the light curves of higher energy bands peak earlier (see Figure~\ref{fig:lcs}). First, we see a short-lived X-ray flare, which quickly decays as more material builds up around the black hole and absorbs the X-ray emission (see Figure~\ref{fig:SEDevolution}). At the same time, the optical and UV light curves begin to rise as the X-ray and EUV emission is reprocessed to longer wavelengths. Notably, it takes much longer than the peak timescale of the fallback rate for half of the mass bound to the black hole to actually return and join the outflowing wind \citep[e.g. see][]{mockler_energy_2021}. As shown in Figure~\ref{fig:SEDevolution}, even at the peak of the optical light curve (which is delayed by a $\sim$~month from the peak of the fallback rate), less than half of the bound mass has returned to the black hole (and so an even smaller fraction is in the wind). 
The UV bands peak slightly before the optical bands -- `UVW2' and `UVM2' are the bluest and peak earliest at 48 and 45 days respectively, while `i' and `r' bands are the reddest and peak latest at 55 days. However, this effect is less noticeable compared to the offset with the X-rays because these bands are much closer to each other in energy than the X-ray bands are to the UV, and therefore have similar light curve shapes (and also much longer peaks). Both early X-ray flares and the general behavior of lower energy bands peaking later has been observed in real events, (most notably in the TDE AT2022dsb/ASASSN-22cs) as well as in the 3D multigroup RHD simulations of \citet[][]{giron_multigroup_2026}. This suggests that detailed UV data (along with optical) during the rising phase of TDEs can probe the formation timescale of the source relative to the mass fallback rate and initial disruption. We discuss additional comparisons of this phenomena with both observations and simulations in Section~\ref{sec:tempevol}. 
We note that choosing other parameters or setups for our simulation would likely produce different results -- for example, if the hot X-ray source turned on later in the simulation, it would encounter a higher optical depth wind from the beginning, and the initial emission would likely be cooler (redder) than what we see here.

The majority of the reprocessed luminosity in our simulations actually comes out at far/extreme UV wavelengths (between $\sim 200-1500 \rm \AA$, see Figure~\ref{fig:SEDevolution}). Because of this, the peak of the bolometric luminosity corresponds to the peak of the extreme UV luminosity. However, the peak of the observed optical and UV luminosity is delayed with respect to the bolometric luminosity. 
This creates a longer/slower rise in the optical and UV luminosity compared to the bolometric luminosity at the earliest times. However, the shape of the optical/UV luminosity curve is very similar to the shape of the bolometric luminosity curve once the luminosity is within $\sim 30\%$ of peak. If you shift the bolometric light curve forward $\sim 20$ days it would approximately line up with the optical/UV light curves (see Figure~\ref{fig:lcs}). 

The fact that the SED peaks at extreme UV wavelengths also means that estimates of the total energy from optical and UV photometry would likely under-predict the true emitted energy by an order of magnitude or more (see blackbody curves that reproduce the optical/UV emission plotted for comparison in Figure~\ref{fig:SEDevolution}). Many previous works have also made this point, and we discuss it in further detail in Section~\ref{sec:disc_euv}.

One caveat here is that we are only including the opacity contributions of oxygen, helium, and hydrogen. Adding more metals may slightly increase the the optical/UV luminosity by increasing the efficiency of redistributing X-ray and EUV photons to lower energies, but we leave this to future work. 

We find that the optical and UV evolution of this particular simulation is reasonably similar to the evolution of AT2020ocn (plotted for comparison in Figure~\ref{fig:lcs}), though we note that we do not try to fit this or any other event in our modeling here. AT2020ocn was a TDE discovered at z=0.07 by the Zwicky Transient Facility (ZTF). The optical/UV light curve of AT2020ocn was presented in \citet{hammerstein_final_2023}), and its black hole mass is estimated to be between $\rm log_{10} M_h \sim 5.8 - 6.7$ (where the lower estimate comes from disk modeling constraints in \citealt{cao_tidal_2024-1} and the higher estimate from the $M-\sigma$ relation, \citealt{hammerstein_integral_2023}). This puts the black hole mass in our model  ($\rm log_{10} M_h = 6.48$) right in the middle of the constraints for this event, and is good confirmation that the time evolution of the light curve is similar to what is observed for an actual TDE in the same black hole mass range. There are a few additional interesting points about the comparison with AT2020ocn, but we save these for later discussion (see Section~\ref{sec:disc_obs}).

\begin{table}
\centering
\hspace{-2em}\begin{tabular}{c|c|c}
    \hline
    band & $\rm log_{\rm 10}(L_{\rm max})$ & $t_{\rm peak}$ \\
    & (ergs/s) & (days) \\
    \hline
    $L_{\rm bol}$ & 44.3 & 28 \\
    UVW2 & 42.2 & 47 \\
    UVM2 & 41.8 & 45 \\
    UVW1 & 41.9 & 50 \\
    g & 41.7 & 49 \\
    r & 41.5 & 55 \\
    i & 41.1 & 55 \\
    X-ray (0.2-1 keV) & 42.3 & 7 \\
    \hline
\end{tabular}
   \caption{The (intrinsic) peak times and maximum luminosities for the bolometric luminosity and each observable band shown in Figure~\ref{fig:lcs}. }
\label{tab:Lmax} 
\end{table}

\subsection{Reprocessing of X-rays}

\begin{figure*}[ht!]
\begin{center}
\includegraphics[width = \textwidth]{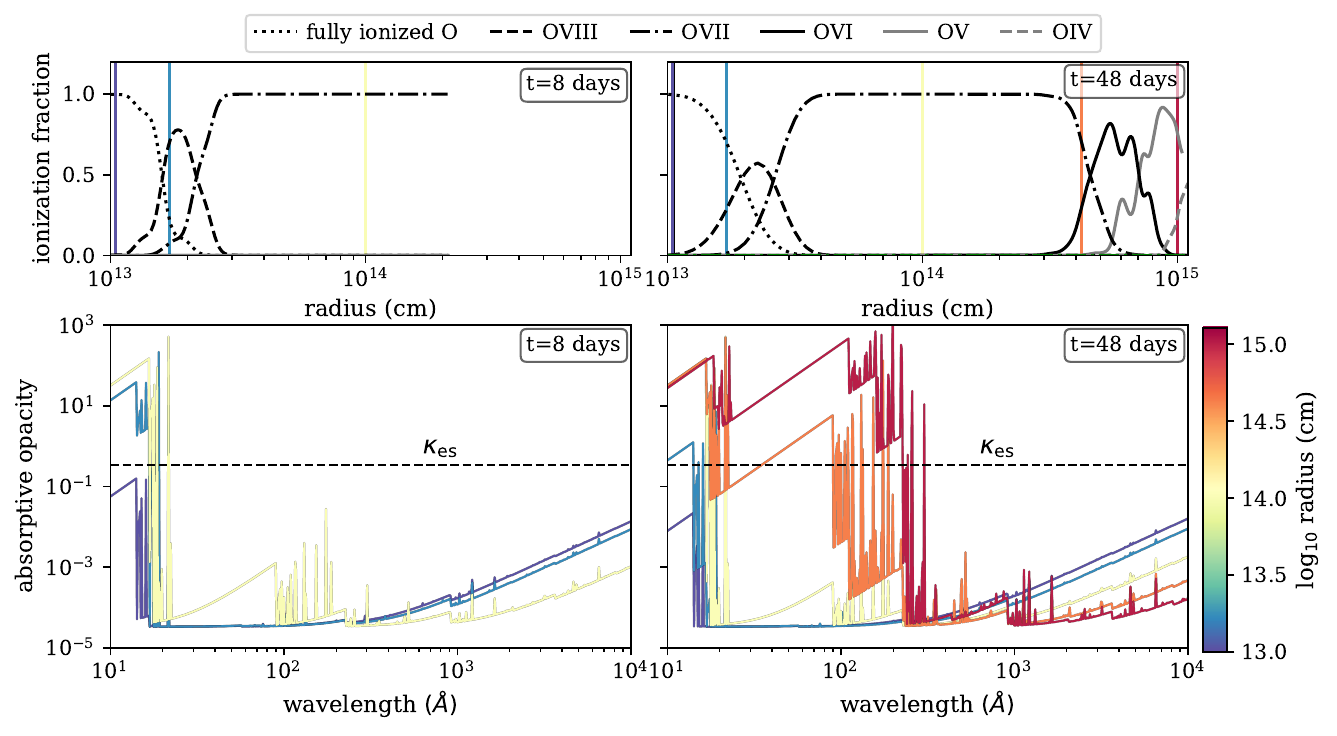} 
\end{center}
\caption{ {\bf Top:} Fraction of oxygen in different ionization states as a function of radius at two representative times in the light curve. Opacities at radii denoted by rainbow vertical lines are plotted in bottom plots. 
{\bf Bottom:} Opacity due to bound-free, free-free, and bound-bound transitions at various radii throughout outflows plotted at the same times as ionization states in top plots. Prominent ionization edges at 17, 90, and 108$\AA$ correspond to the ionization of OVII, OVI, and OV respectively. The ionization edge between 226-228$\AA$ is due to the ionization of both OIV and HeII. The ionization edge at 910$\AA$ is from HI and OI. The upward slope of the opacity redward of $1000\AA$ comes from free-free emission.} Near the peak of the optical/UV light curve, the increased fraction of oxygen in lower ionization states at large radii (orange and red lines in right plots) results in the X-ray emission getting absorbed.
\label{fig:ionization+opacity}

\end{figure*}

We find the bound-free (or photoionization) opacity contribution from oxygen to be very efficient at absorbing the soft X-ray radiation in our simulations. This is similar to past work on TDEs \citep[][]{roth_x-ray_2016-1, parkinson_optical_2022, thomsen_dynamical_2022-1}, as well as to work on warm absorbers in AGN  \citep[][]{reynolds_warm_1995, blustin_nature_2005, mizumoto_thermally_2019}.
In the first few days of our simulation, before very much material has been launched in the wind, the majority of the X-ray radiation escapes (see top panels of Figure~\ref{fig:lcs} and Figure~\ref{fig:SEDevolution}). At these times, the gas is highly ionized and OVII is the dominant species of oxygen over the majority of the simulation\footnote{This is because there is a much larger jump in ionization energy between OVII and OVIII than between OVI and OVII, and so OVII is the dominant ion over a much larger range in temperature.}.
However, very quickly, as more material builds up in the wind, the gas in the outer half is able to cool enough so that lower ionization states of oxygen appear and can absorb X-rays efficiently through photoionization (see Figure~\ref{fig:ionization+opacity}).
Notably, at any given time, the wavelength-dependent opacity varies dramatically at different radii throughout the outflow. For example, at 48 days (near optical/UV peak), the photoionization edges in the opacity plot between $\sim 20-100 \rm \AA$ correspond to the presence of OVI and lower ionization states, and only show up at radii of $\gtrsim 3 \times 10^{14}$cm. At smaller radii within the outflow (or at polar viewing angles in a 3D simulation), soft X-rays in this wavelength range would still be visible (as shown in the bottom plots in Figure~\ref{fig:lcs}).

Before the bolometric peak, the source luminosity increases with time and the source SED has more and more energy in the X-ray part of the spectrum. However, because our wind is also increasing at the same time, reprocessing by the gas wins out and the amount of observed X-ray emission decreases with time (see Figure~\ref{fig:SEDevolution}). This is also reflected in the top panel of Figure~\ref{fig:ionization+opacity}, which shows that some of the oxygen in the outflow remains in lower ionization states (OIV, OV, and OVI) at t=48 days, resulting in high opacities to X-ray emission. %

This effect is separate from the ``viewing angle'' effect often discussed in the reprocessing of TDE disk emission \citep[e.g.][]{dai_unified_2018}. Here we are simulating a reprocessing scenario analogous to looking at the forming disk from an intermediate or ``side-on'' angle in work such as \citet{thomsen_dynamical_2022-1} (as opposed to an angle near the poles). The fraction of the X-ray radiation is changing because of the {\it time-dependence} of the outflow and the reprocessing layer, which, for most viewing angles, will likely become more optically thick near peak as the mass fallback rate and mass outflow rate increase. The influence of a time-dependent optical depth from outflows on the X-ray to optical/UV ratio is also described in recent 3D work \citep[][]{huang_x-ray_2025, giron_multigroup_2026}, and we discuss how the 3D work compares to the simulations shown here in Section~\ref{sec:disc_xray}.

\subsection{Opacities}\label{sec:opacities}

In Figure~\ref{fig:SEDevolution}, we explore the importance of including different opacities (electron scattering, free-free, bound-free, bound-bound) in our simulation. We find that while runs with only electron scattering and bound-free opacities produce SEDs with similar overall shapes, reproducing the optical tail over the full light curve requires including both free-free and bound-bound transitions. In fact, we find that post-peak, including line opacities (bound-bound transitions) increases the optical/UV continuum by up to a factor of $\sim 5$ (see Figure~\ref{fig:SEDevolution}). 
Because of the importance of these transitions to the optical luminosity, the ionization state of the gas did affect the relative amount of optical luminosity across runs that did or did not include free-free and/or bound-bound opacities, with larger differences at later times (post-bolometric peak), when the gas was less ionized. 

We also find that the wavelength-dependent opacity varied dramatically with both radius in the outflow and with time. This is not surprising, given the orders of magnitude changes in the mass flow rates and source luminosity, however in Figure~\ref{fig:ionization+opacity} we can compare directly between the opacity early on the rise and near the UV/optical peak. We also also show how dramatically it varies with radius at each of these times.

For example, we find that the continuum opacity to optical photons is dominated by free-free emission, which is highest here at the smallest radii. Free-free opacity scales as $\kappa_{\rm ff} \propto \rho T^{-3.5}$, and we find that the higher density at small radii wins out over the hotter temperature (though this could in principle change with e.g. a lower mass black hole and an even hotter source).
We also find that the opacity to the softest X-rays is actually highest at large radii, where the fraction of oxygen in lower ionization states is the highest -- hinting that this particular wavelength range is likely very time-dependent in addition to viewing angle dependent as the outflows surrounding the disk evolve with time and the ionization state of the gas changes. 

As described above, to include bound-bound transitions we use a simplified treatment of line absorption with a line absorption efficiency parameter `$\epsilon$' which can vary from $\epsilon =0$ (no absorption) to $\epsilon =1$ (100\% absorption). Based on \citet[]{Kasen:2006a} and the relatively featureless spectra characteristic of TDEs (compared to supernovae), we chose $\epsilon = 0.1$. However, we find that for $\epsilon$ values between 0.01-0.3 the shape of the resulting SEDs in the observable (UV and optical) bands do not change significantly (see Appendix~\ref{sec:lineabsorption}). Therefore a wide range of line absorption efficiencies produces similar optical and UV luminosities.\footnote{We find that varying this parameter does not change the SED continuum emission at optical/UV wavelengths significantly until $\epsilon$ nears 1, at which point it is clear that the lines become much stronger than is typically observed for TDEs (see Appendix~\ref{sec:nLTE_comparison}). However, even for $\epsilon = 1$, the SEDs \textit{near peak} are very similar across runs, as the gas is highly ionized.}. We also checked individual epochs of our simulations against 
post-processed non-local thermodynamic equilibrium (nLTE) calculations using {\tt Sedona} \citep{thomsen_dynamical_2022-1}, and find similar results for the amount of emission reprocessed to optical and UV wavelengths under similar gas and temperature conditions (see Appendix~\ref{sec:nLTE_comparison}). 

While the focus of this work is not to reproduce TDE line profiles (and we only include H, He, and O in these models), we note that in our runs with bound-bound transitions included, we do resolve the H$\alpha$ and He II 4686 lines observed in most TDEs, as shown in Figure~\ref{fig:SEDevolution}\footnote{See \citet[]{aspegren_emission_2026} for a detailed analysis of the effect of luminosity and gas conditions on the relative strength of H and He lines in UV-bright transients.}.

\subsection{Energetics}\label{sec:energetics}

Connecting the observed emission in TDEs to the energetics of the inner source is critical for constraining the efficiency of accretion disk formation and super-Eddington accretion.  Calculating the total emitted luminosity for observed events is difficult for multiple reasons -- first of all, as described above, it is often quite difficult to constrain the escaping bolometric luminosity with just the optical/UV data. 
In addition to the bolometric corrections required to correctly calculate the total escaping luminosity, significant energy will be lost through adiabatic expansion and some can also go into radiative acceleration and the ionization of the gas. For the luminosities and mass flow rates in our simulations (and more generally for most TDEs), we expect the dominant effect on the overall energy budget to be adiabatic losses from luminosity trapped in the wind. These adiabatic losses and their effect on the output bolometric luminosity can also be approximated analytically. 
Here we show how analytical approximations of both the bolometric energy evolution and the photosphere evolution compare to our simulations to better understand the energy budget and energy losses through the wind.

\begin{figure}[ht!]
\begin{center}
\includegraphics[width = \columnwidth]{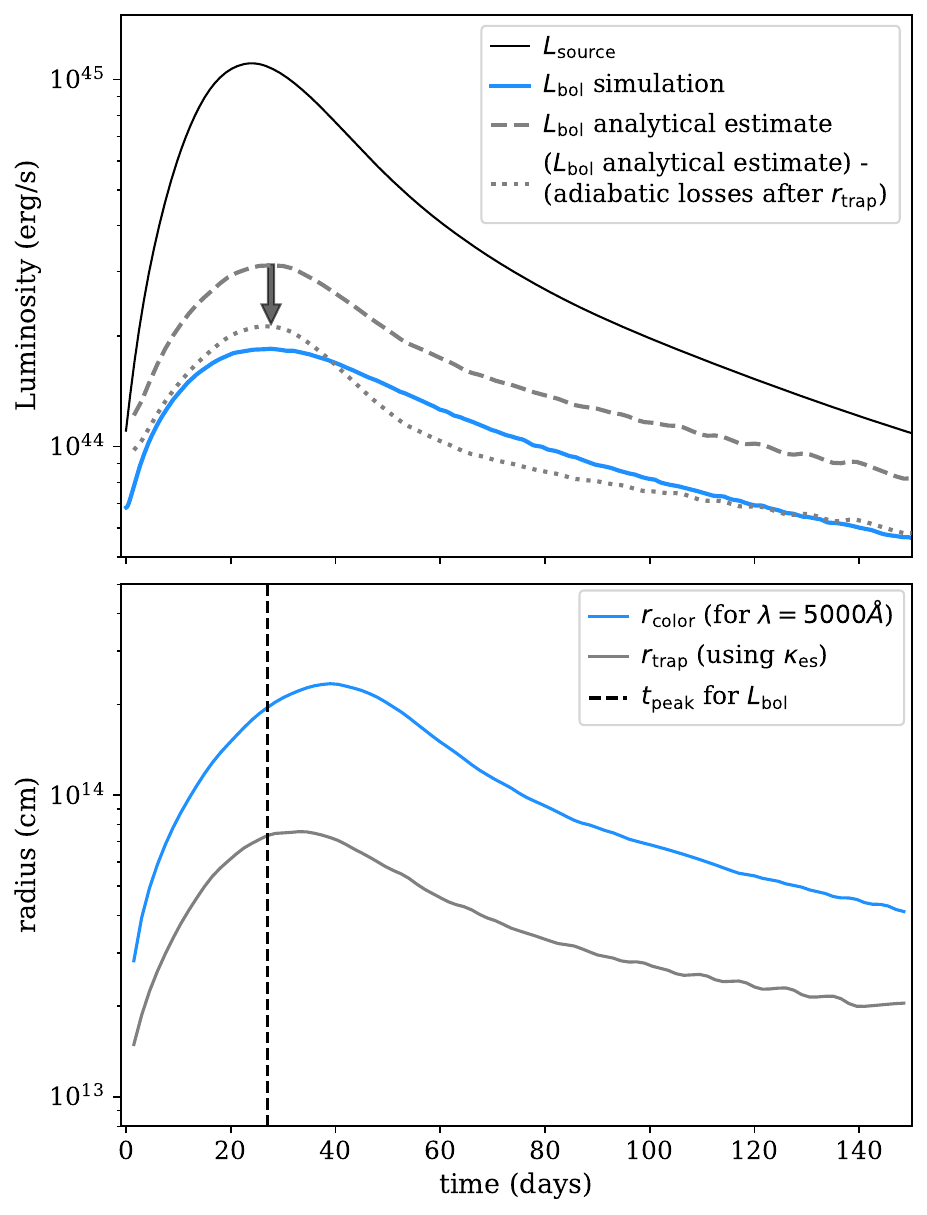} 
\end{center}
\caption{ {\bf Top:} We find that adiabatic losses result in a peak luminosity that is $\sim 6 \times$ lower than the input luminosity ($L_{\rm source}$). Analytic predictions based on the evolving density profile in the simulations and $L_{\rm source}$ predict a higher luminosity (smaller adiabatic losses, dashed line). However, these predictions assume that all adiabatic losses occur interior to $r_{\rm trap}$ (that all luminosity interior to $R_{\rm trap}$ is advected, and all luminosity outside $r_{\rm trap}$ diffuses). In reality, this transition is more gradual. If we subtract the adiabatic losses from outside the trapping radius (from the simulation) from the analytical prediction of the luminosity (dotted line), we find that it is much closer to the simulation output luminosity. {\bf Bottom:} The trapping radius and (optical) color radius, compared to the time of peak optical luminosity. 
The color radius ($r_{\rm color}$) is wavelength dependent, so here we set $\kappa_{\rm abs}(t) = \kappa_{\rm abs}(t)$ at $\lambda = 5000 \rm \AA$ as we are interested in the reprocessing of optical emission. The optical color radius continues to grow after the bolometric luminosity peaks, similar to what we see from the optical luminosity. 
\label{fig:energetics}
}
\end{figure}

We find qualitatively similar results for the relation between our source luminosity and our bolometric luminosity to other simulation work that models wind reprocessing \citep[e.g.][]{roth_what_2018, calderon_moving-mesh_2021}. As expected, the dominant effect on the bolometric luminosity curve are the adiabatic losses from luminosity trapped in the wind. We can approximate this analytically by assuming that luminosity is advected within the trapping radius and diffuses outside of it, so that the trapping radius sets the bolometric luminosity that makes it out to the observer. Here we define the trapping radius as where $t_{\rm diff}(r_{\rm trap}) = t_{\rm adv}(r_{\rm trap})$, and follow \citet[]{piro_wind-reprocessed_2020} by approximating  $t_{\rm diff} (r)= \frac{\tau(r)}{c}\frac{(r_w - r)r}{r_w}$ and $t_{\rm adv} = (r - r_{\rm in})/v_w$ for the purposes of the analytical approximation, which matches the expected limits a $r <<r_w$ and $r\sim r_w$. 
For the analytical calculation we also use $\kappa_{\rm es}$ to calculate $\tau(r)$, as this is the dominant opacity near the trapping radius, and using it allows for simpler comparisons with past work. 
Given this, we can estimate the luminosity that escapes by scaling the luminosity at $r_{\rm in}$ as follows:
\begin{equation}
    L[r_{\rm trap}, t] = L[r_{\rm in}, t - t_{\rm adv}] \Big(\frac{\rho[r_{\rm trap}, t]}{\rho_{\rm in}[t]}\Big)^{4/3}\Big(\frac{r_{\rm trap}[t]}{r_{\rm in}}\Big)^{2}, 
\end{equation}
Where it is assumed that within the trapping radius, the gas is supported by radiation pressure and so $T \propto \rho^{1/3}$ and therefore $L \propto r^2 T^4 \propto r^2\rho^{4/3}$ \citep[e.g.][]{Strubbe:2009a}. We also have included the time-delay for luminosity to advect to $r_{\rm trap}$. This calculation is plotted as the dashed line in Figure~\ref{fig:energetics}.
            
While the analytical result provides a reasonable order of magnitude estimate, we do note some differences from the simulation. The most important difference is that the amount of adiabatic losses in our simulation is about a factor of two higher than predicted by the analytical estimate based on the size of the trapping radius (see Figure~\ref{fig:energetics}). The reason for this discrepancy appears to be largely due to the fact that the trapping radius is generally defined analytically as where $t_{\rm diff}(r_{\rm trap}) = t_{\rm adv} (r_{\rm trap})$, and so at the estimated trapping radius, about half of the luminosity is diffusing and about half is advected.
Therefore, there are additional adiabatic losses after the trapping radius -- not all of the luminosity at that radius makes it out of the wind. To check this, we calculate these additional losses outside of the trapping radius in our simulation by integrating:

\begin{equation}
\begin{split}
    L_{\rm ad} &= \int_{r_{\rm trap}}^{r_{\rm out}} p_{\rm rad} (\nabla \cdot \mathbf{v}) dV \\
    & = \int_{r_{\rm trap}}^{r_{\rm out}} \frac{1}{3} E_{\rm rad}[r] \Big(\frac{dv}{dr} + 2 \frac{v[r]}{r}\Big) 4 \pi r^2 dr,
\end{split}
\end{equation}

Where this is no longer a purely analytical calculation as we take $E_{\rm rad}$, $v(r)$, and $\frac{dv}{dr}$ from the simulation. We find that including these additional adiabatic losses outside of $r_{\rm trap}$ brings the analytical prediction much closer to the luminosity curve we get from the simulation (see dotted line in Figure~\ref{fig:energetics} for a calculation including these additional losses). The difference from the semi-analytic prediction is particularly noticeable near the peak of the mass fallback rate and light curve, when the optical depth is highest and luminosity is trapped longest. 
We note that despite significant energy going into adiabatic work, it is still much less than the initial kinetic energy of the wind, and so the wind is accelerated by $<1\%$ of its initial velocity in the simulation.
\footnote{Note that we initialize the gas with $v_{\rm wind}=0.01c$, if we chose a higher $L_{\rm source}$ or lower initial velocity we would expect a larger percent change in wind velocity.} 

Intriguingly, we only find small delays in the bolometric luminosity peak timescale when comparing to the source luminosity peak timescale from diffusion/advection through the gas (the peak of the bolometric luminosity shows a $\sim 4$ day delay from the peak of the source luminosity, see blue and black solid lines in Figure~\ref{fig:energetics}). The delay in the peak timescale in our simulation is comparable to what we get analytically when only including the advection delay (defined as the time it takes the wind and the photons trapped within to reach the trapping radius = $r_{\rm trap}/v_{\rm wind}$). This is the delay between the dotted and dashed gray lines and the black line in Figure~\ref{fig:energetics} ($\sim 3$ days).
However, this means the delay we get in the simulation is shorter than what would be predicted analytically using the advection timescale to the trapping radius plus the diffusion time out through the rest of the wind. 
We note again that we use $\kappa_{\rm es}$ for these analytical calculations -- if we instead used the Rosseland mean opacity, this could only act to increase the analytical diffusion timescale, as $\kappa_{\rm es}$ is the minimum value in this simulation (and a good approximation of the opacity near the trapping radius). 

Finally, we note that we would expect less adiabatic losses in a multi-dimensional simulation that allowed radiation to travel out preferentially through less dense material. As explained earlier (and from comparison with 3D work), we expect the observed SED and spectra would look similar to 3D simulations viewed away from the poles \citep[e.g. in][]{thomsen_dynamical_2022-1}.

In addition to calculating the analytical trapping radius to estimate adiabatic losses through the gas, we also plot the color radius evolution, where we have defined the color radius ($r_{\rm color}$) as where the effective optical depth to absorption for a given wavelength equals 1. This effectively defines a photosphere radius for that wavelength:
\begin{equation}
    \tau_{\rm eff} = \int_{r_{\rm color}}^{r_w}{ (3 \kappa_{\rm abs} \kappa_{\rm es})^{1/2} \rho dr} = 1, 
\end{equation}
Because the value of $\kappa_{\rm abs}(t)$ is wavelength dependent, the color radius changes with wavelength, and in Figure~\ref{fig:energetics} we plot the color radius at $\lambda = 5000 \rm \AA$ \footnote{Note that this is nearly identical to the color radius in g band calculated using the weighted average of the absorptive opacity over the Palomar g band filter given its transmission curve.}. 

The color radius provides a more accurate approximation of the observed photosphere surface (again, for a given wavelength) as, depending on the optical depth, photons can continue to exchange energy with the surrounding gas outside of the electron scattering trapping radius (see bottom plot in Figure~\ref{fig:energetics}). Notably, the color radius in Figure~\ref{fig:energetics} (at $\lambda = 5000 \rm \AA$) reaches its maximum after the trapping radius and after $L_{\rm bol}$, emphasizing that the optical photosphere is still being built up after the peak of the bolometric luminosity. 

\section{Discussion}\label{sec:discussion}

Here we discuss in more detail the implications of our results as well as what they mean for TDE observations. 

\subsection{Temperature evolution \& pre-peak cooling}\label{sec:tempevol}
One of the main takeaways of this work is the delay observed between the peak of the bolometric luminosity and the peak of the observable optical/UV luminosity. This delay comes from the time it takes to build up an optically thick reprocessing layer around the black hole, and is therefore dependent both on the rate of mass return to the black hole, the fraction of mass ejected to large radii, as well as the source luminosity timescale. For example, we find that when we estimate an optical `color' radius (using the simulation's opacity at $5000 \rm \AA$), the peak of this radius occurs after the peak of the trapping radius and the bolometric luminosity and closer to when the optical luminosity peaks (see Figure~\ref{fig:energetics}). 

This `pre-peak cooling' during the optical rise that is described above and shown in Figure~\ref{fig:lcs} is dependent on the source luminosity convolved with the rate of ejected mass. Notably, there are visible delays between the individual optical and UV bands in addition to the much larger delay between the bolometric luminosity curve and the average optical/UV luminosity curve (see Figure~\ref{fig:ionization+opacity} and Table~\ref{tab:Lmax}). This behavior has also been seen observationally. For example, \citet[][]{hinkle_discovery_2021} note they find a 17 day delay between UVW2 and `i' band for AT2019azh (ASASSN-19dj). Smaller delays are also visible in other TDEs with detailed pre-peak UV data such as AT2018dyb \citep[][]{leloudas_spectral_2019,Holoien:2020},  AT2019ahk \citep[ASASSN-19bt, the first TDE discovered with TESS,][]{Holoien:2019}, and  AT2020zso \citep[][]{hammerstein_final_2023-1}. While it is rare to have X-ray coverage in the month before optical peak, as X-ray follow-up is generally triggered after the TDE is confirmed in optical, in \citet[][]{malyali_transient_2024}, the authors report that an X-ray flare was serendipitously discovered by eROSITA right before an optical TDE was detected by ASASSN (AT2022dsb/ASASSN-22cs). It has also been shown that many TDEs brighten in X-ray at late times \citep[][]{gezari_x-ray_2017,holoien_unusual_2018, wevers_evidence_2019,hinkle_discovery_2021, guolo_systematic_2023}, as the optical flare fades, which may also be consistent with the reprocessing scenario, as the fallback rate decreases and the material surrounding the black hole becomes optically thin at late times (the inverse of what we see at very early times in our simulation).

In this work we use a single setup as a proof of concept that wind reprocessing of an X-ray and EUV bright source can reproduce the observed optical/UV evolution in a tidal disruption event. However, future work modeling a variety of different source luminosity and mass outflow evolutions will be necessary to determine how these parameters affect observables. For example, it is likely that the fraction of mass ejected at early times is dependent on how efficient the initial stream collisions are at redistributing energy and angular momentum \citep[and therefore on the orbital properties and thermodynamic properties of the initial streams,][]{bonnerot_first_2021,jankovic_spin-induced_2024, andalman_resolving_2025}. It is also likely that the fraction of mass ejected near peak is dependent on how super-Eddington the fallback rate is. 

Recent work by \citet[][]{huang_x-ray_2025} find that the fraction of the fallback rate ejected in outflows increases on the rise of the TDE as the fallback rate becomes more super-Eddington.
This can change the temperature evolution of the flare-- for example the timescale of the `pre-peak cooling' described above, as well as whether the optical/UV temperature remains relatively constant after optical peak or continues to evolve. In this work, the optical/UV temperature increases pre-peak, but appears $\sim$constant after peak (see Figure~\ref{fig:lcs}). However, in \citet[][]{huang_x-ray_2025}, the optical/UV temperature was nearly constant pre-peak. This could be partly due to the fact that the mass fallback rate used in \citet[][]{huang_x-ray_2025} is scaled to a lower mass black hole ($1 \times 10^6 M_\odot$) compared to what was used in this work ($3 \times 10^6 M_\odot$), and therefore starts at a higher value and increases more rapidly than what we use here. This means there is more material surrounding the hot source at early times in \citet[][]{huang_x-ray_2025} compared to this work. In a more extreme example, \citet[][]{kremer_wind-reprocessed_2023}, produced semi-analytical modeling of wind-reprocessing of stellar mass black hole TDEs. These were even more super-Eddington and the very high mass ejection rates were shown to produce strong temperature evolution (cooling) in the observable SED -- similar to the observed `luminous fast-coolers' \citep[][]{nicholl_at2022aedm_2023}. This was because the optical depth built up high enough that the effective `color' radius for optical and UV luminosities continued growing long after the peak of the bolometric luminosity.

\subsection{EUV emission and missing energy}\label{sec:disc_euv}
We show here that the SEDs from our simulations peak in the EUV over the full evolution of the light curve, leading to under-predictions of the emitted energy by about a factor of 10 when looking at only the optical and UV luminosities (see blackbody curves in Figure~\ref{fig:SEDevolution}).

Similar behavior has been shown before in steady-state or post-processing radiation transport simulations \citep[][]{roth_x-ray_2016-1, thomsen_dynamical_2022-1, parkinson_optical_2022}, and recently in time-dependent RHD multi-group simulations \citep[][]{giron_multigroup_2026}. It has been used to explain the `missing energy' problem in TDEs \citep[where their observed optical and UV luminosity is much lower than would be predicted for the bolometric luminosity from an accreting black hole fed near $\dot{M}_{\rm edd}$, e.g.,][]{lu_missing_2018}. It is also consistent with the systematically higher total emitted energy calculated for TDEs with observational constraints on dust-reprocessed IR emission \citep[][]{van_velzen_reverberation_2021,  masterson_new_2024}. 

The extreme UV emission predicted in our simulations would also potentially be observable through dust reprocessing. If there is dust in the nuclear region, the (otherwise unobservable) extreme UV emission will heat it up, and cause it to emit in the infrared at later times. 
Many TDEs have now been detected at late times in infrared emission \citep[][]{van_velzen_discovery_2016}, as have ambiguous nuclear flares \citep[at least some of which may be particularly extreme TDEs][]{hinkle_mid-infrared_2024}, though this is still a small subset of the total number of optical/UV TDEs \citep[][]{jiang_infrared_2021}. There is also population of infrared-selected TDE candidates of which most have no optical counterparts \citep[][]{masterson_new_2024},
though in a more general selection of mid-infrared nuclear outbursts, the majority do not seem to be consistent with TDE origins \citep[][]{dodd_mid-infrared_2023}.

In this work we also show how the time-dependent \textit{efficiency} of reprocessing can produce an observable time-lag in the X-ray to optical (and even UV to redder optical) light curves. It is possible this could be used to help constrain the source temperature and luminosity and therefore extent of missing energy. This motivates high cadence UV observations near the peak of the light curve, when UV emission may start declining even as e.g. r-band emission is $\sim$flat or rising. 

\subsection{X-ray observability}\label{sec:disc_xray}
X-rays are observable at early times in our simulation, however they are quickly absorbed as the mass outflow increases around the source. This time-dependent mass outflow produces a similar effect to the `viewing angle' dependence first described in \citet[][]{dai_unified_2018} and explored further in \citet[][]{thomsen_dynamical_2022-1}, where viewing the TDE near the poles versus near the equatorial plane of the disk leads to vastly different optical depths and X-ray-to-optical luminosity ratios. However, it is inherently a completely different effect, and in fact both the viewing angle dependence and the effect of the time-dependent mass outflow should be occurring together in a real TDE. Our simulation, which is in 1D, is likely a reasonable approximation of intermediate viewing angles in a real TDE (see Figure~\ref{fig:thomsen_comparison} for a comparison with a steady-state 3D simulation). 

We note that changes in the mass outflow and source luminosity evolutions could also change the X-ray evolution. Recent 3D RHD simulations by both \citet[][]{huang_x-ray_2025} and \citet[][]{giron_multigroup_2026}  modeling the initial disk formation process and light curve rise also show interesting X-ray variability. While they differ in details, both find that the X-ray to optical/UV luminosity ratio varies significantly pre-peak due to changes in the outflows surrounding the forming disk. \citet[][]{huang_x-ray_2025} also find that viewing angles through the forming disk (away from the poles) are initially X-ray bright, but then the X-ray emission drops off as the gas surrounding the black hole increases with time, only to re-emerge when radiation pressure driven outflows clear out a cavity. In general \citet[][]{huang_x-ray_2025} find higher levels of X-ray to optical variability than our 1D simulations as outflows clear some viewing angles to shocks and the forming disk but obscure others. Meanwhile, \citet[][]{giron_multigroup_2026} predict similar X-ray flare behavior to what we find at early times, with emission that peaks and decays as the optical/UV emission continues to rise. In their simulation this is due to early circularizing shocks (specifically the nozzle shock) that produce X-rays at small radii before the optical depth of the surrounding gas grows large enough to obscure it.

We find that most of the X-rays in our simulation are absorbed at the outer edge of the wind (see bottom panel of Figure~\ref{fig:lcs}), as the wind's opacity to X-rays is highly dependent on the ionization state of metals (in our case oxygen), and the metals will be the least ionized at the outer edges\footnote{We note that adding additionally heavier metals (i.e. iron-group elements) could impact where exactly in the outflow  X-rays are absorbed. However, \citet[][]{huang_x-ray_2025} found similar behavior for X-ray absorption as a function of radius in a TDE with similar gas conditions using TOPS opacities that include iron group elements.} (see Figure~\ref{fig:ionization+opacity}). This means that even in cases where no X-rays are observed, almost the entire envelope can still be irradiated by X-ray photons. Because of this, we predict that lines excited by X-ray photons may sometimes be observed even with no detected X-ray continuum, and that this could happen at many different viewing angles as it does not require X-rays to ever fully escape. However, we caution readers that while we trust the general X-ray evolution in our simulations, the exact shape of our output SED at X-ray wavelengths is an approximation, as it is dependent on our input SED, which is a blackbody (this is not the case for the UV/optical portion of the SED which is fully reprocessed). In future simulations we hope to use source SEDs informed by e.g. hydrodynamic simulations of the forming disk. 

\subsection{Comparison to AT2020ocn}\label{sec:disc_obs}

As we show in Figure~\ref{fig:lcs}, the optical/UV evolution of our simulation reproduces the behavior of AT2020ocn reasonably well. AT2020ocn is a TDE from the ZTF survey with an estimated black hole mass between $\rm log_{10} M_h \sim 5.8 - 6.7$ \citep[][]{cao_tidal_2024, hammerstein_integral_2023}, consistent with our simulation parameters (which use $\rm log_{10} M_h = 6.48$).
There are a few additional interesting points about the comparison with AT2020ocn.

First, this TDE shows X-ray flaring behavior \citep{cao_tidal_2024-1}. If the optical and UV luminosity from this event is from wind reprocessing from a disk, this would imply that asymmetries in the reprocessing region/viewing angle effects are necessary to explain both the X-ray and optical/UV luminosity \citep[similar to what is seen in][]{huang_x-ray_2025}. Our simulation is in 1D, and so cannot produce these asymmetries, however this motivates further study either in 3D or with varying mass outflow rates in 1D to approximate different lines of sight.

Another point of interest is that previous blackbody fits to the optical and UV data of AT2020ocn found best fit temperatures that peaked at UV wavelengths and significant host extinction to match both the optical and UV data \citep[E(B-V) $ = 0.76^{+0.14}_{-0.44}$][]{hammerstein_final_2023}. However, we find that with our wind reprocessing model, our light curves 
can match the observed optical/UV emission without adding heavy host extinction (and instead with SEDs that peak in the EUV). X-ray modeling in \citep{cao_tidal_2024-1} for AT2020ocn also found negligible host extinction, emphasizing the uncertainty in modeling the optical and UV SEDs of TDEs as blackbodies without strong constraints on the peak of the SED or the far UV emission.

\section{Summary}\label{sec:summary}

In this work we simulate the reprocessing of emission in a tidal disruption event using monte carlo radiation transport and 1D moving-mesh hydrodynamics over the rise, peak, and initial decay of the light curve. We find that an outflowing (though not unbound) wind consistent with those seen in 3D hydro simulations \citep[e.g.][]{steinberg_origins_2022, price_eddington_2024, huang_x-ray_2025} is able to successfully reprocess the X-ray and EUV-bright source emission to optical and UV wavelengths. We also find that this emission has similar luminosity evolution and optical/UV colors to observed TDEs (see Figure~\ref{fig:lcs}). 

However, we also find that there is the time-delay between the bolometric and optical/UV luminosity curves that comes from the time it takes to build up the reprocessing layer around the hot, X-ray bright source. Perhaps surprisingly, this does not significantly alter the actual \textit{shape} of the curve in our simulation, however it does shift it forwards in time (by $\sim 20$ days, see top panel of Figure~\ref{fig:lcs}).  Emission from the forming disk (at size scales comparable to the tidal radius) is predicted to be bright at times $\lesssim t_{\rm peak, fb}$ \citep[e.g.][]{steinberg_origins_2022}, but we show here that there may not be sufficient material around at those times to produce a strong \textit{optical} flare. 
This time-dependent reprocessing also produces a $\sim 20$ day X-ray flare at early times, which declines as fallback material obscures the hot source and the optical/UV luminosity rises. This implies that TDEs that circularize promptly may be X-ray bright before they are discovered in optical surveys. This is similar to the pre-peak flare caught serendipitously with eROSITA before an optical/UV TDE in \citet[]{malyali_transient_2024}.

Because this is a 1D simulation, material entirely obscures the inner source, and it is comparable to results from 3D simulations observed at viewing angles far from the poles. Future work is planned to compare with other viewing angles by adjusting the density of the outflow based on comparisons with 3D simulations. 

\subsection{Key Results}

Here we briefly describe the main results of this work:

\begin{itemize}
    \item The wind outflow is able to reprocess X-ray and EUV bright source emission to optical/UV wavelengths, and produces a light curve similar to observed events (see Figure~\ref{fig:lcs}).
    \item Bluer bands peak earlier. This is most apparent when comparing the observed optical/UV light curve to the bolometric curve (which most closely traces the EUV emission), but is also visible in the delay between the UV and redder optical bands. The optical/UV light curves have a similar shape to the bolometric curve \textit{but} lag the bolometric by $3-4$ weeks -- the time it takes for sufficient material to build up to reprocess the source emission (see Figure~\ref{fig:lcs} and Table~\ref{tab:Lmax}).
    \item X-rays are absorbed at the outer edge of the wind, and therefore most of the wind is irradiated by X-rays photons even when the X-ray continuum is not observable (see bottom panel of Figure~\ref{fig:lcs}).
    \item The opacity changes dramatically over the course of the simulation as the ionization states of the gas change (see Figure~\ref{fig:ionization+opacity}). Before optical peak, bound-free and free-free opacities only slightly underestimate the continuum opacity and do a reasonable job of reproducing the observed optical/UV continuum (though they cannot reproduce the EUV continuum). However, after optical peak, line opacities increase the optical/UV continuum by up to factors of 5 (see Figure~\ref{fig:SEDevolution}). 
    \item Adiabatic losses reduce the input luminosity by a factor of $\sim 6$ near peak (less at earlier and later times, see Figure~\ref{fig:energetics}). This is more than predicted from analytic approximations, likely due to the fact that those generally utilize a definition of the trapping radius as where $L_{\rm adv} \approx L_{\rm diff}$, and so half the luminosity is still advecting outside of this radius and suffers further adiabatic losses before it escapes.
\end{itemize}

\subsection{Next Steps}

These simulations are a proof of concept showing that our setup of time-dependent wind-reprocessing of an evolving X-ray and EUV bright source can successfully reproduce light curves of observed tidal disruption events. We plan to build on this with a detailed parameter study that will address how the luminosity, mass flow, outflow velocity, and other properties affect the efficiency of reprocessing and observed light curves.

\section*{Acknowledgments}
This work was performed in part at Aspen Center for Physics, which is supported by National Science Foundation grant PHY-2210452. It was also supported in part by grant NSF PHY-2309135 to the Kavli Institute for Theoretical Physics (KITP). This work was performed under the auspices of the U.S. Department of Energy by Lawrence Livermore National Laboratory under Contract DE-AC52-07NA27344. This research used resources of the National Energy Research Scientific Computing Center (NERSC), a U.S. Department of Energy Office of Science User Facility located at Lawrence Berkeley National Laboratory, operated under Contract No. DE-AC02-05CH11231 using NERSC award NP-ERCAP-0025048. B.M. is grateful for support from the Carnegie Theoretical Astrophysics Center.  

\vspace{5mm}

\software{astropy \citep{Astropy-Collaboration:2013a} 
          }
\appendix


\section{Hydrodynamic evolution}\label{sec:hydrodynamics}

\begin{figure}[ht!]
\begin{center}
\includegraphics[width = 0.9\textwidth]{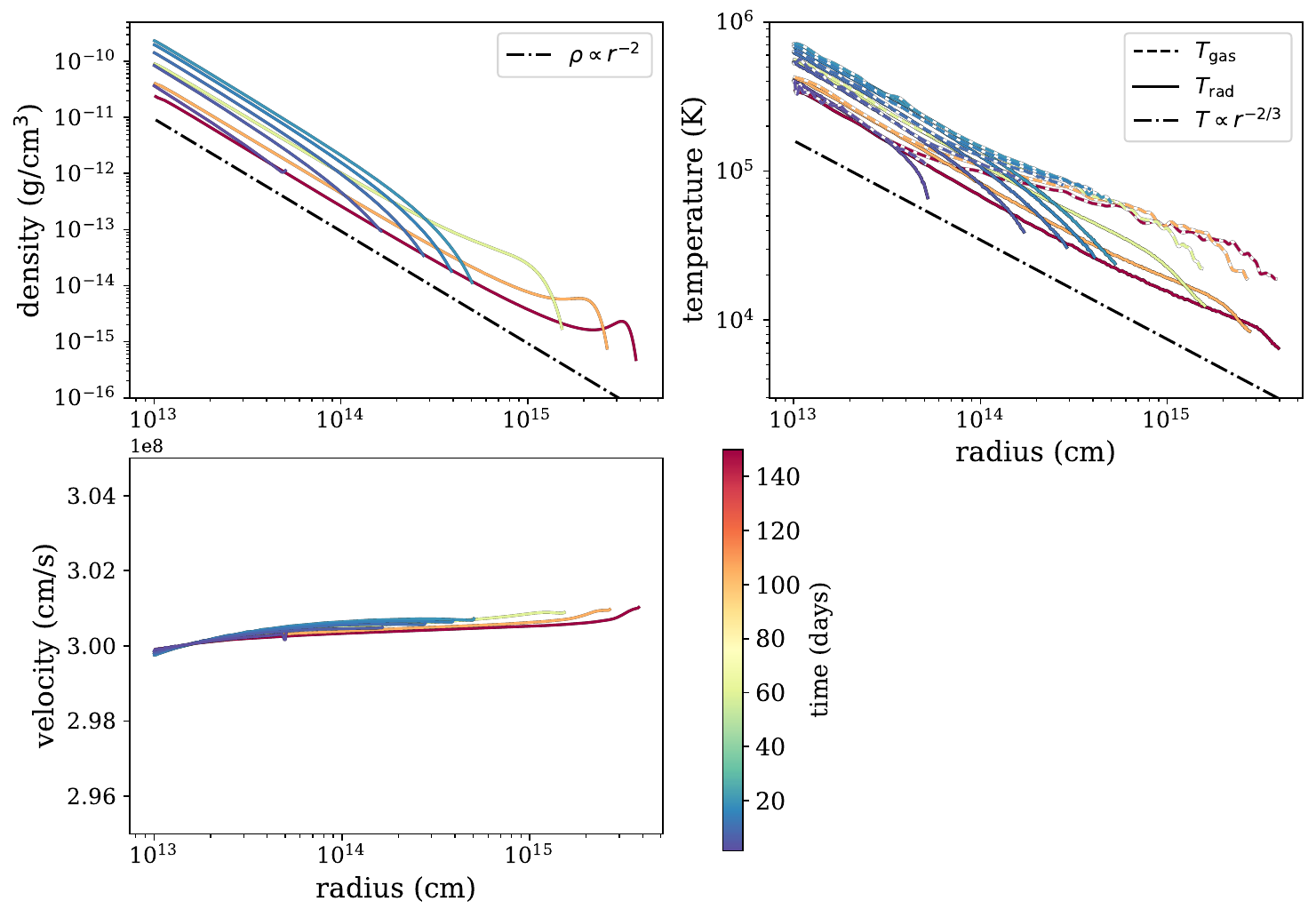} 
\end{center}
\caption{ Relevant hydro quantities as a function of time and radius from the fiducial simulation run.
\label{fig:hydro}
}
\end{figure}

We have included plots of the hydrodynamic evolution below (Fig~\ref{fig:hydro}). We input material and source luminosity at the inner boundary ($10^{13}$ cm). The source luminosity evolution is plotted in Figure~\ref{fig:energetics}, and the source $\dot{M}$ is from the hydrodynamic simulation of a disruption of a $\gamma = 4/3$ polytrope $1 M_\odot$ star scaled to a $3\times 10^{6} M_\odot$ black hole (from \citealt{guillochon_hydrodynamical_2013}, available through the {\tt MOSFiT} TDE model described in \citealt{mockler_weighing_2019}). It is input with a velocity of $v = 0.01$c.

For this initial velocity and the luminosities in our simulation, the material experiences only very slight acceleration from radiation pressure (increasing the velocity by $< 1\%$ over the course of the simulation).  At early times, $\dot{M}$ is increasing with time, and so the density profile initially falls off more steeply than the $r^{-2}$ profile expected for constant $\dot{M}$ and velocity. However, because $\dot{M}$ reaches a peak and turns over, there is a bump in $\dot{M}$ in the outer regions at later times. Note that while the inner region appears to approach an $r^{-2}$ profile, it is always slightly shallower as $\dot{M} \propto t^{-5/3}$ at late times.  

The temperature of the gas and radiation is in equilibrium up until around the trapping radius, at which point the gas is no longer able to cool as efficiently as the radiation and the two diverge. This is important for accurately calculating the ionization state of the gas, and therefore its opacity, which determines the optical and UV light curve that we observe. We can also compare the energy in gas and radiation by approximating the radiation temperature as $T_{\rm rad} = (e_{\rm rad}/a)^{1/4}$ (where $a$ is the radiation constant) and comparing it to the gas temperature. Given that the radiation field is not a blackbody, this is not the same as the electron temperature, and the radiation field at a given radius will not peak at this temperature when the gas and radiation are not in thermal equilibrium. However, it is still a useful comparison to the gas temperature as it is a direct measure of the energy in radiation. Using this definition, we find that the radiation temperature in the inner regions approximately follows the prediction for an adiabatically expanding wind \citep[$T \propto \rho^{1/3} \propto r^{-2/3}$ for $\rho \propto r^{-2}$, e.g.,][]{Strubbe:2009a}.

\section{Treatment of line absorption}\label{sec:lineabsorption}

We find that when varying the value of of the line absorption parameter $\epsilon$, the magnitude of the continuum emission does not vary significantly {\it except} in the extreme UV. We do see that the strength of the lines varies (as expected), but the continuum emission in the observable optical, UV, and X-ray bands varies by less than 10\%. We note that when the H$\alpha$ and He II lines lie within the `r' and `g' bands, this can change the relative optical luminosity in those bands by up to a factor of $\sim 2$. Given this, we choose the intermediate value of $\epsilon = 0.1$ for our fiducial simulations, which (unlike the $\epsilon =0.01$ simulation), produces hydrogen and helium lines similar to those observed in many TDEs. However, this is a conservative estimate of the `g' and `r' band peak luminosity, which is slightly higher in the simulation with $\epsilon= 0.3$. We additionally ran a test with $\epsilon= 1.0$ but found that the line strengths of HeII and H$\alpha$ were much higher than what we see observationally. However, even with $\epsilon= 1.0$, the continuum changed by less than a factor of 2. Finally, the range in epsilon explored here and our fiducial choice of $\epsilon = 0.1$ are also motivated by predictions from \citet[][]{roth_x-ray_2016-1}, who analytically estimated a value of $\epsilon = 0.06$ for TDE envelopes with similar conditions to what we model in this work.

\begin{figure}[ht!]
\begin{center}
\includegraphics[width = 0.6\columnwidth]{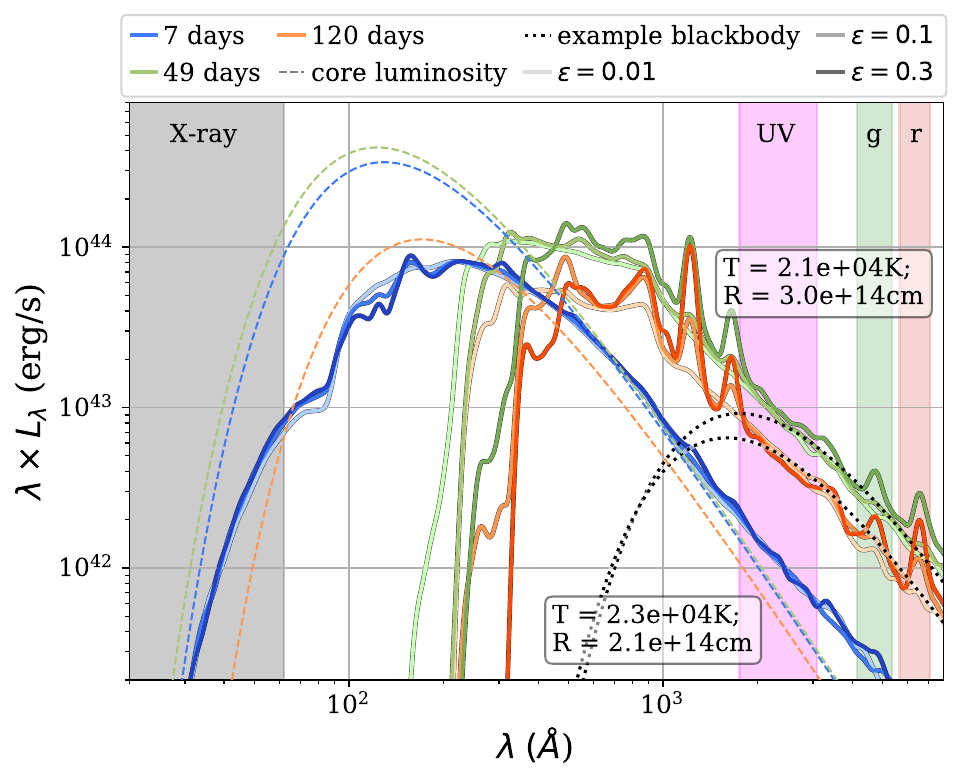} 
\end{center}
\caption{ Here we compare the SED evolution for simulations with three different values of the line opacity $\epsilon$ parameter: 0.01, 0.1, and 0.3. 
\label{fig:epsiloncomparison}
}
\end{figure}

\section{Comparison with non-LTE models}\label{sec:nLTE_comparison}

Here we compare individual epochs of our simulation with \citet[]{thomsen_dynamical_2022-1}, who use {\tt Sedona} to post-process different viewing angles of steady-state simulations of TDE-like disks using non-LTE radiation transport. To make it easier to compare between the simulations, we use a gaussian smoothing kernel to approximate the lower resolution in the \citet[]{thomsen_dynamical_2022-1} runs. These simulations are of a $10^6 M_\odot$ black hole and have accretion rates of 7, 12, and 24 $\dot{M}_{\rm edd}$, comparable to the Eddington ratios in our simulation (for a $3 \times 10^6 M_\odot$ black hole). The peak fallback rate in our simulation is $26 \dot{M}_{\rm edd}$ (assuming the canonical thin disk efficiency of 0.1), however at optical peak the fallback rate is $14 \dot{M}_{\rm edd}$. Given this, we compare to the \citet[]{thomsen_dynamical_2022-1} with $12 \dot{M}_{\rm edd}$. We note that because the outflows in our simulation are time-dependent, the optical depth is significantly lower than for a similar time-independent run with the same peak mass fallback rate (providing further motivation to compare with the $12 \dot{M}_{\rm edd}$ \citealt[]{thomsen_dynamical_2022-1} simulation, which is close to the time-averaged mass fallback rate in our simulation for all mass that has returned by optical peak). Our simulation is also spherically symmetric, and so the optical depth is lower than it would be at edge-on viewing angles from a puffy disk simulation with a similar overall amount of mass, but higher than it would be for polar viewing angles. Keeping these caveats in mind, we do find that our SED at early times (blue line) is similar to the intermediate viewing angle ($50$ degrees) from \citet[]{thomsen_dynamical_2022-1}, and our simulation's SED near optical peak (orange line) is similar to the near edge-on ($70$ degree) viewing angle from the same work. This implies that even without the full non-LTE calculation, we are able to reproduce similar reprocessed SEDs. However, for precise calculations of line profiles, non-LTE calculations are likely required.

\begin{figure}[ht!]
\begin{center}
\includegraphics[width = 0.6\columnwidth]{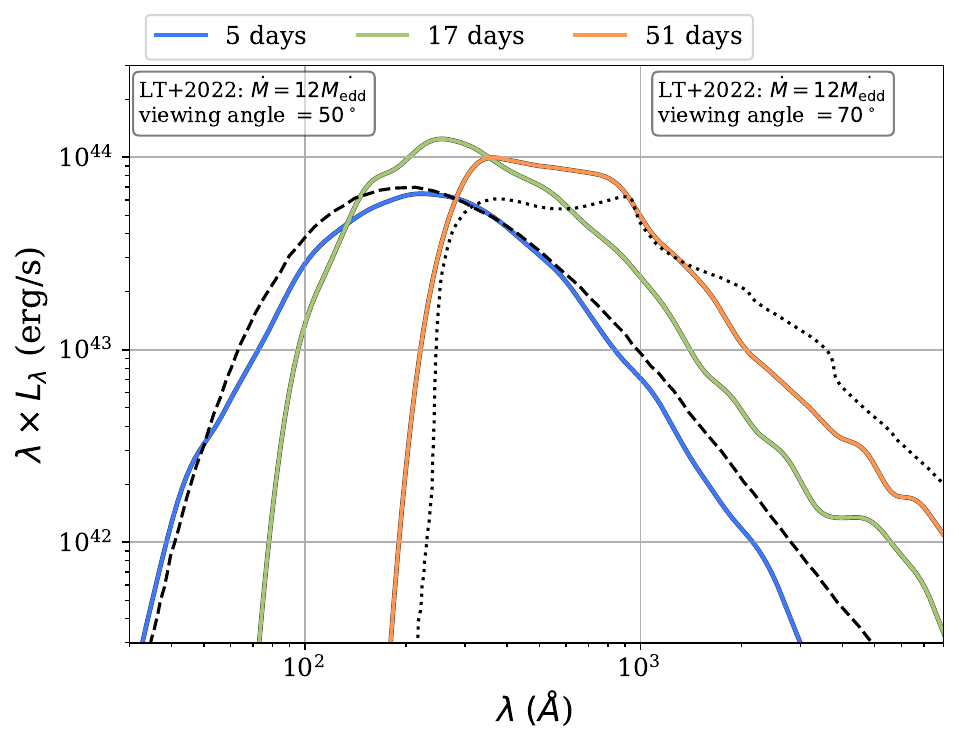} 
\end{center}
\caption{We compare 3 epochs of our simulation with {\tt Sedona} non-LTE post-processing results from \citet[]{thomsen_dynamical_2022-1}. We use a gaussian smoothing kernel to better compare with the lower resolution in \citet[]{thomsen_dynamical_2022-1}.
\label{fig:thomsen_comparison}
}
\end{figure}

\section{Test of radiation hydrodynamics}\label{sec:radhydrotest}

\begin{figure}[ht!]
\begin{center}
\includegraphics[width = 0.6\columnwidth]{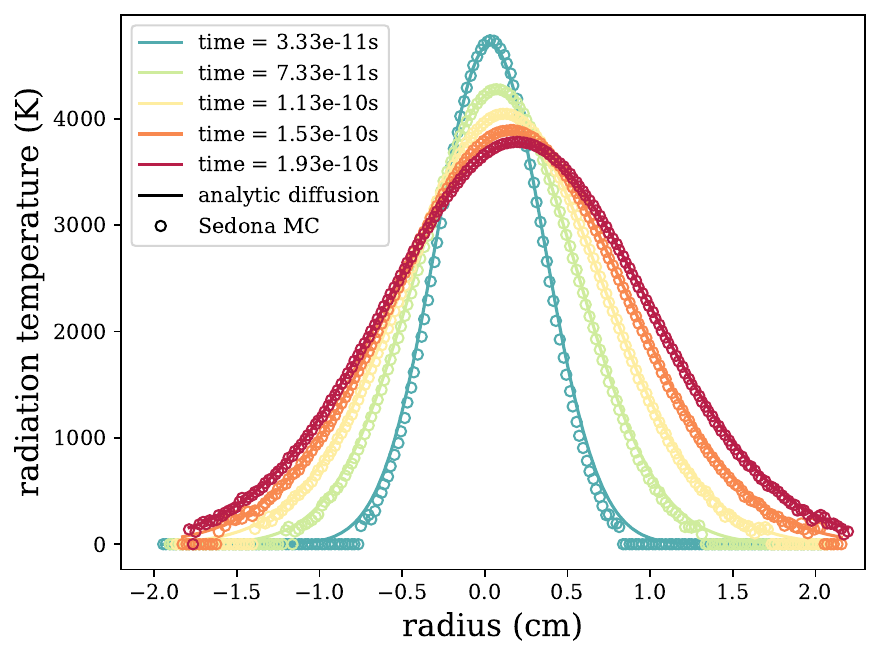} 
\end{center}
\caption{The advected pulse problem solved analytically in the diffusion approximation (solid lines) and using {\tt Sedona} (open circles). 
\label{fig:advected_pulse}
}
\end{figure}

The {\tt Sedona} implementation used here is described in detail in \citet[][]{khatami_physics_2024}. Some tests of the transport can be found in that work, as well as in the appendix of \citet[][]{khatami_landscape_2024}. Here we include our implementation of the advected pulse problem, which tests both the Monte Carlo radiation transport and hydrodynamics in the code as photons are advected along with the gas. We solve it assuming gray opacity and pure scattering ($\epsilon = 0$). We use 200 moving mesh zones that are initialized with zone sizes of 0.02cm. Our timesteps are allowed to vary between $10^{-13}$s and $10^{-11}$s, and we run the simulation for $2 \times 10^{-10}$s. We initialize the pulse in the central zone with $10^7$ photon packets and a radiation temperature of $10^4$K, consistent with a radiation energy density of 75.7 ergs/cm$^3$. We compare it with the analytical solution solved in the diffusion approximation in Figure~\ref{fig:advected_pulse}. The details of the analytical solution can be found in \citet[][]{harries_algorithm_2011, Noebauer:2012a}.

\bibliography{zotero_library, library, sample631}{}

@article{Holoien:2020,
	adsurl = {https://ui.adsabs.harvard.edu/abs/2020ApJ...898..161H},
	archiveprefix = {arXiv},
	author = {{Holoien}, Thomas W. -S. and {Auchettl}, Katie and {Tucker}, Michael A. and {Shappee}, Benjamin J. and {Patel}, Shannon G. and {Miller-Jones}, James C.~A. and {Mockler}, Brenna and {Groenewald}, Dani{\`e}l N. and {Hinkle}, Jason T. and {Brown}, Jonathan S. and {Kochanek}, Christopher S. and {Stanek}, K.~Z. and {Chen}, Ping and {Dong}, Subo and {Prieto}, Jose L. and {Thompson}, Todd A. and {Beaton}, Rachael L. and {Connor}, Thomas and {Cowperthwaite}, Philip S. and {Dahmen}, Linnea and {French}, K. Decker and {Morrell}, Nidia and {Buckley}, David A.~H. and {Gromadzki}, Mariusz and {Roy}, Rupak and {Coulter}, David A. and {Dimitriadis}, Georgios and {Foley}, Ryan J. and {Kilpatrick}, Charles D. and {Piro}, Anthony L. and {Rojas-Bravo}, C{\'e}sar and {Siebert}, Matthew R. and {Velzen}, Sjoert van},
	doi = {10.3847/1538-4357/ab9f3d},
	eid = {161},
	eprint = {2003.13693},
	journal = {\apj},
	month = aug,
	number = {2},
	pages = {161},
	primaryclass = {astro-ph.HE},
	title = {{The Rise and Fall of ASASSN-18pg: Following a TDE from Early to Late Times}},
	volume = {898},
	year = 2020}

@article{Holoien:2019,
	adsurl = {https://ui.adsabs.harvard.edu/abs/2019ApJ...883..111H},
	archiveprefix = {arXiv},
	author = {{Holoien}, Thomas W. -S. and {Vallely}, Patrick J. and {Auchettl}, Katie and {Stanek}, K.~Z. and {Kochanek}, Christopher S. and {French}, K. Decker and {Prieto}, Jose L. and {Shappee}, Benjamin J. and {Brown}, Jonathan S. and {Fausnaugh}, Michael M. and {Dong}, Subo and {Thompson}, Todd A. and {Bose}, Subhash and {Neustadt}, Jack M.~M. and {Cacella}, P. and {Brimacombe}, J. and {Kendurkar}, Malhar R. and {Beaton}, Rachael L. and {Boutsia}, Konstantina and {Chomiuk}, Laura and {Connor}, Thomas and {Morrell}, Nidia and {Newman}, Andrew B. and {Rudie}, Gwen C. and {Shishkovksy}, Laura and {Strader}, Jay},
	doi = {10.3847/1538-4357/ab3c66},
	eid = {111},
	eprint = {1904.09293},
	journal = {\apj},
	month = oct,
	number = {2},
	pages = {111},
	primaryclass = {astro-ph.HE},
	title = {{Discovery and Early Evolution of ASASSN-19bt, the First TDE Detected by TESS}},
	volume = {883},
	year = 2019}

@article{Gallegos-Garcia:2018,
	adsurl = {https://ui.adsabs.harvard.edu/abs/2018ApJ...857..109G},
	archiveprefix = {arXiv},
	author = {{Gallegos-Garcia}, Monica and {Law-Smith}, Jamie and {Ramirez-Ruiz}, Enrico},
	doi = {10.3847/1538-4357/aab5b8},
	eid = {109},
	eprint = {1801.03497},
	journal = {\apj},
	month = apr,
	number = {2},
	pages = {109},
	primaryclass = {astro-ph.HE},
	title = {{Tidal Disruptions of Main-sequence Stars of Varying Mass and Age: Inferences from the Composition of the Fallback Material}},
	volume = {857},
	year = 2018}

@article{Leloudas:2019,
	adsurl = {https://ui.adsabs.harvard.edu/abs/2019ApJ...887..218L},
	archiveprefix = {arXiv},
	author = {{Leloudas}, Giorgos and {Dai}, Lixin and {Arcavi}, Iair and {Vreeswijk}, Paul M. and {Mockler}, Brenna and {Roy}, Rupak and {Malesani}, Daniele B. and {Schulze}, Steve and {Wevers}, Thomas and {Fraser}, Morgan and {Ramirez-Ruiz}, Enrico and {Auchettl}, Katie and {Burke}, Jamison and {Cannizzaro}, Giacomo and {Charalampopoulos}, Panos and {Chen}, Ting-Wan and {Cikota}, Aleksand ar and {Della Valle}, Massimo and {Galbany}, Lluis and {Gromadzki}, Mariusz and {Heintz}, Kasper E. and {Hiramatsu}, Daichi and {Jonker}, Peter G. and {Kostrzewa-Rutkowska}, Zuzanna and {Maguire}, Kate and {Mandel}, Ilya and {Nicholl}, Matt and {Onori}, Francesca and {Roth}, Nathaniel and {Smartt}, Stephen J. and {Wyrzykowski}, Lukasz and {Young}, Dave R.},
	doi = {10.3847/1538-4357/ab5792},
	eid = {218},
	eprint = {1903.03120},
	journal = {\apj},
	month = dec,
	number = {2},
	pages = {218},
	primaryclass = {astro-ph.HE},
	title = {{The Spectral Evolution of AT 2018dyb and the Presence of Metal Lines in Tidal Disruption Events}},
	volume = {887},
	year = 2019}

@article{Dai:2018,
	adsurl = {https://ui.adsabs.harvard.edu/abs/2018ApJ...859L..20D},
	archiveprefix = {arXiv},
	author = {{Dai}, Lixin and {McKinney}, Jonathan C. and {Roth}, Nathaniel and {Ramirez-Ruiz}, Enrico and {Miller}, M. Coleman},
	doi = {10.3847/2041-8213/aab429},
	eid = {L20},
	eprint = {1803.03265},
	journal = {\apjl},
	month = jun,
	number = {2},
	pages = {L20},
	primaryclass = {astro-ph.HE},
	title = {{A Unified Model for Tidal Disruption Events}},
	volume = {859},
	year = 2018}

@article{Guillochon:2018,
	adsurl = {https://ui.adsabs.harvard.edu/abs/2018ApJS..236....6G},
	archiveprefix = {arXiv},
	author = {{Guillochon}, James and {Nicholl}, Matt and {Villar}, V. Ashley and {Mockler}, Brenna and {Narayan}, Gautham and {Mandel}, Kaisey S. and {Berger}, Edo and {Williams}, Peter K.~G.},
	doi = {10.3847/1538-4365/aab761},
	eid = {6},
	eprint = {1710.02145},
	journal = {\apjs},
	month = may,
	number = {1},
	pages = {6},
	primaryclass = {astro-ph.IM},
	title = {{MOSFiT: Modular Open Source Fitter for Transients}},
	volume = {236},
	year = 2018}

@article{Astropy-Collaboration:2013a,
	adsurl = {http://adsabs.harvard.edu/abs/2013A%26A...558A..33A},
	archiveprefix = {arXiv},
	arxivurl = {http://arxiv.org/abs/1307.6212},
	author = {{Astropy Collaboration} and {Robitaille}, T.~P. and {Tollerud}, E.~J. and {Greenfield}, P. and {Droettboom}, M. and {Bray}, E. and {Aldcroft}, T. and {Davis}, M. and {Ginsburg}, A. and {Price-Whelan}, A.~M. and {Kerzendorf}, W.~E. and {Conley}, A. and {Crighton}, N. and {Barbary}, K. and {Muna}, D. and {Ferguson}, H. and {Grollier}, F. and {Parikh}, M.~M. and {Nair}, P.~H. and {Unther}, H.~M. and {Deil}, C. and {Woillez}, J. and {Conseil}, S. and {Kramer}, R. and {Turner}, J.~E.~H. and {Singer}, L. and {Fox}, R. and {Weaver}, B.~A. and {Zabalza}, V. and {Edwards}, Z.~I. and {Azalee Bostroem}, K. and {Burke}, D.~J. and {Casey}, A.~R. and {Crawford}, S.~M. and {Dencheva}, N. and {Ely}, J. and {Jenness}, T. and {Labrie}, K. and {Lim}, P.~L. and {Pierfederici}, F. and {Pontzen}, A. and {Ptak}, A. and {Refsdal}, B. and {Servillat}, M. and {Streicher}, O.},
	doi = {10.1051/0004-6361/201322068},
	eid = {A33},
	eprint = {1307.6212},
	journal = {\aap},
	month = oct,
	pages = {A33},
	primaryclass = {astro-ph.IM},
	title = {{Astropy: A community Python package for astronomy}},
	volume = 558,
	year = 2013}

@article{Auchettl:2017b,
	adscomment = {10 pages, 6 figures. To be submitted to ApJ},
	adsurl = {http://adsabs.harvard.edu/abs/2017arXiv170306141A},
	archiveprefix = {arXiv},
	arxivurl = {http://arxiv.org/abs/1703.06141},
	author = {{Auchettl}, K. and {Ramirez-Ruiz}, E. and {Guillochon}, J.},
	eprint = {1703.06141},
	journal = {ArXiv e-prints},
	month = mar,
	primaryclass = {astro-ph.HE},
	title = {{Comparison of the X-ray emission from Tidal Disruption Events with those of Active Galactic Nuclei}},
	year = 2017}

@article{Auchettl:2017a,
	adsurl = {http://adsabs.harvard.edu/abs/2017ApJ...838..149A},
	archiveprefix = {arXiv},
	arxivurl = {http://arxiv.org/abs/1611.02291},
	author = {{Auchettl}, K. and {Guillochon}, J. and {Ramirez-Ruiz}, E.},
	doi = {10.3847/1538-4357/aa633b},
	eid = {149},
	eprint = {1611.02291},
	jornal = {ArXiv e-prints},
	journal = {\apj},
	month = apr,
	pages = {149},
	primaryclass = {astro-ph.HE},
	title = {{New Physical Insights about Tidal Disruption Events from a Comprehensive Observational Inventory at X-Ray Wavelengths}},
	volume = 838,
	year = 2017}

@article{Roth:2016a,
	adsurl = {http://adsabs.harvard.edu/abs/2016ApJ...827....3R},
	archiveprefix = {arXiv},
	arxivurl = {http://arxiv.org/abs/1510.08454},
	author = {{Roth}, N. and {Kasen}, D. and {Guillochon}, J. and {Ramirez-Ruiz}, E.},
	doi = {10.3847/0004-637X/827/1/3},
	eid = {3},
	eprint = {1510.08454},
	journal = {\apj},
	month = aug,
	pages = {3},
	primaryclass = {astro-ph.HE},
	title = {{The X-Ray through Optical Fluxes and Line Strengths of Tidal Disruption Events}},
	volume = 827,
	year = 2016}

@article{van-Velzen:2011a,
	adsurl = {http://adsabs.harvard.edu/abs/2011ApJ...741...73V},
	archiveprefix = {arXiv},
	arxivurl = {http://arXiv.org/abs/1009.1627},
	author = {{van Velzen}, S. and {Farrar}, G.~R. and {Gezari}, S. and {Morrell}, N. and {Zaritsky}, D. and {{\"O}stman}, L. and {Smith}, M. and {Gelfand}, J. and {Drake}, A.~J.},
	date = {2011/00/01},
	doi = {10.1088/0004-637X/741/2/73},
	eid = {73},
	eprint = {1009.1627},
	journal = {\apj},
	m3 = {10.1088/0004-637X/741/2/73},
	month = nov,
	n2 = {Using archival Sloan Digital Sky Survey (SDSS) multi-epoch imaging data (Stripe 82), we have searched for the tidal disruption of stars by supermassive black holes in non-active galaxies. Two candidate tidal disruption events (TDEs) are identified. The TDE flares have optical blackbody temperatures of 2 ×104 K and observed peak luminosities of Mg = -18.3 and -20.4 (νL ν= 5 ×1042, 4 ×1043 erg s-1, in the rest frame); their cooling rates are very low, qualitatively consistent with expectations for tidal disruption flares. The properties of the TDE candidates are examined using (1) SDSS imaging to compare them to other flares observed in the search, (2) UV emission measured by GALEX, and (3) spectra of the hosts and of one of the flares. Our pipeline excludes optically identifiable AGN hosts, and our variability monitoring over nine years provides strong evidence that these are not flares in hidden AGNs. The spectra and color evolution of the flares are unlike any SN observed to date, their strong late-time UV emission is particularly distinctive, and they are nuclear at high resolution arguing against these being first cases of a previously unobserved class of SNe or more extreme examples of known SN types. Taken together, the observed properties are difficult to reconcile with an SN or an AGN-flare explanation, although an entirely new process specific to the inner few hundred parsecs of non-active galaxies cannot be excluded. Based on our observed rate, we infer that hundreds or thousands of TDEs will be present in current and next-generation optical synoptic surveys. Using the approach outlined here, a TDE candidate sample with O(1) purity can be selected using geometric resolution and host and flare color alone, demonstrating that a campaign to create a large sample of TDEs, with immediate and detailed multi-wavelength follow-up, is feasible. A by-product of this work is quantification of the power spectrum of extreme flares in AGNs.},
	number = {2},
	pages = {73},
	primaryclass = {astro-ph.CO},
	title = {{Optical Discovery of Probable Stellar Tidal Disruption Flares}},
	ty = {JOUR},
	url = {http://adsabs.harvard.edu/cgi-bin/nph-data_query?bibcode=2011ApJ...741...73V&link_type=ABSTRACT},
	volume = 741,
	year = 2011}

@article{Holoien:2014a,
	adsurl = {http://adsabs.harvard.edu/abs/2014MNRAS.445.3263H},
	archiveprefix = {arXiv},
	arxivurl = {http://arXiv.org/abs/1405.1417},
	author = {{Holoien}, T.~W.-S. and {Prieto}, J.~L. and {Bersier}, D. and {Kochanek}, C.~S. and {Stanek}, K.~Z. and {Shappee}, B.~J. and {Grupe}, D. and {Basu}, U. and {Beacom}, J.~F. and {Brimacombe}, J. and {Brown}, J.~S. and {Davis}, A.~B. and {Jencson}, J. and {Pojmanski}, G. and {Szczygie{\l}}, D.~M.},
	date = {2014/00/06},
	doi = {10.1093/mnras/stu1922},
	eprint = {1405.1417},
	journal = {\mnras},
	month = dec,
	pages = {3263-3277},
	publisher = {Cornell University Library},
	title = {{ASASSN-14ae: a tidal disruption event at 200 Mpc}},
	ty = {JOUR},
	url = {http://arxiv.org/abs/1405.1417v1},
	volume = 445,
	year = 2014}

@article{Noebauer:2012a,
	adsurl = {http://adsabs.harvard.edu/abs/2012MNRAS.425.1430N},
	archiveprefix = {arXiv},
	arxivurl = {http://arXiv.org/abs/1206.6263},
	author = {{Noebauer}, U.~M. and {Sim}, S.~A. and {Kromer}, M. and {R{\"o}pke}, F.~K. and {Hillebrandt}, W.},
	doi = {10.1111/j.1365-2966.2012.21600.x},
	eprint = {1206.6263},
	journal = {\mnras},
	month = sep,
	pages = {1430-1444},
	primaryclass = {astro-ph.HE},
	title = {{Monte Carlo radiation hydrodynamics: methods, tests and application to Type Ia supernova ejecta}},
	volume = 425,
	year = 2012}

@article{Kasen:2006a,
	address = {AA(Department of Physics and Astronomy, Johns Hopkins University, 3400 North Charles Street, Baltimore, MD 21218.; Allan C. Davis Fellow.; Space Telescope Science Institute, 3700 San Martin Drive, Baltimore, MD 21218.), AB(Lawrence Berkeley National Laboratory, 1 Cyclotron Road, Berkeley, CA 94720.), AC(Lawrence Berkeley National Laboratory, 1 Cyclotron Road, Berkeley, CA 94720.)},
	author = {Kasen, Daniel and Thomas, R C and Nugent, P},
	date = {2006/00/01},
	journal = {\apj},
	m3 = {10.1086/506190},
	month = {00},
	n2 = {We discuss Monte Carlo techniques for addressing the three-dimensional time-dependent radiative transfer problem in rapidly expanding supernova atmospheres. The transfer code SEDONA has been developed to calculate the light curves, spectra, and polarization of aspherical supernova models. From the onset of free expansion in the supernova ejecta, SEDONA solves the radiative transfer problem self-consistently, including a detailed treatment of gamma-ray transfer from radioactive decay and with a radiative equilibrium solution of the temperature structure. Line fluorescence processes can also be treated directly. No free parameters need be adjusted in the radiative transfer calculation, providing a direct link between multidimensional hydrodynamic explosion models and observations. We describe the computational techniques applied in SEDONA and verify the code by comparison to existing calculations. We find that convergence of the Monte Carlo method is rapid and stable even for complicated multidimensional configurations. We also investigate the accuracy of a few commonly applied approximations in supernova transfer, namely, the stationarity approximation and the two-level atom expansion opacity formalism.},
	pages = {366},
	title = {Time-dependent Monte Carlo Radiative Transfer Calculations for Three-dimensional Supernova Spectra, Light Curves, and Polarization},
	ty = {JOUR},
	url = {http://adsabs.harvard.edu/cgi-bin/nph-data_query?bibcode=2006ApJ...651..366K&link_type=ABSTRACT},
	volume = {651},
	year = {2006}}

@article{Rees:1988a,
	address = {AA(Cambridge University, England)},
	author = {Rees, Martin J},
	date = {1988/00/01},
	journal = {\nat},
	m3 = {10.1038/333523a0},
	month = {00},
	n2 = {Stars in galactic nuclei can be captured or tidally disrupted by a central black hole. Some debris would be ejected at high speed; the remainder would be swallowed by the hole, causing a bright flare lasting at most a few years. Such phenomena are compatible with the presence of 10 to the 6th-10 to the 8th solar mass holes in the nuclei of many nearby galaxies. Stellar disruption may have interesting consequences in our own Galactic Center if an approximately 10 to the 6th solar mass hole lurks there.},
	pages = {523},
	publisher = {Cornell University Library},
	title = {Tidal disruption of stars by black holes of 10 to the 6th-10 to the 8th solar masses in nearby galaxies},
	ty = {JOUR},
	url = {http://adsabs.harvard.edu/abs/1988Natur.333..523R},
	volume = {333},
	year = {1988}}

@article{Strubbe:2009a,
	address = {AA(Astronomy Department and Theoretical Astrophysics Center, 601 Campbell Hall, University of California, Berkeley, CA 94720, USA; ), AB(Astronomy Department and Theoretical Astrophysics Center, 601 Campbell Hall, University of California, Berkeley, CA 94720, USA; )},
	author = {Strubbe, Linda E and Quataert, Eliot},
	date = {2009/00/01},
	journal = {\mnras},
	m3 = {10.1111/j.1365-2966.2009.15599.x},
	month = {00},
	n2 = {A star that wanders too close to a massive black hole (BH) is shredded by the BH's tidal gravity. Stellar gas falls back to the BH at a rate initially exceeding the Eddington rate, releasing a flare of energy. In anticipation of upcoming transient surveys, we predict the light curves and spectra of tidal flares as a function of time, highlighting the unique signatures of tidal flares at optical and near-infrared wavelengths. A reasonable fraction of the gas initially bound to the BH is likely blown away when the fallback rate is super-Eddington at early times. This outflow produces an optical luminosity comparable to that of a supernova; such events have durations of \~{}10 d and may have been missed in supernova searches that exclude the nuclear regions of galaxies. When the fallback rate subsides below Eddington, the gas accretes onto the BH via a thin disc whose emission peaks in the ultraviolet to soft X-rays. Some of this emission is reprocessed by the unbound stellar debris, producing a spectrum of very broad emission lines (with no corresponding narrow forbidden lines). These lines are the strongest for BHs with MBH \~{} 105-106Msolar and thus optical surveys are particularly sensitive to the lowest mass BHs in galactic nuclei. Calibrating our models to ROSAT and Galaxy Evolution Explorer (GALEX) observations, we predict detection rates for Panoramic Survey Telescope and Rapid Response System (Pan-STARRS), the Palomar Transient Factory (PTF) and the Large Synoptic Survey Telescope (LSST) and highlight some of the observational challenges associated with studying tidal disruption events in the optical. Upcoming surveys such as Pan-STARRS should detect at least several events per year, and may detect many more if current models of outflows during super-Eddington accretion are reasonably accurate. These surveys will significantly improve our knowledge of stellar dynamics in galactic nuclei, the physics of super-Eddington accretion, the demography of intermediate mass BHs and the role of tidal disruption in the growth of massive BHs.},
	pages = {2070},
	title = {Optical flares from the tidal disruption of stars by massive black holes},
	ty = {JOUR},
	url = {http://adsabs.harvard.edu/cgi-bin/nph-data_query?bibcode=2009MNRAS.400.2070S&link_type=ABSTRACT},
	volume = {400},
	year = {2009}}

@article{bonnerot_first_2021,
	title = {First light from tidal disruption events},
	volume = {504},
	issn = {0035-8711},
	url = {https://ui.adsabs.harvard.edu/abs/2021MNRAS.504.4885B},
	doi = {10.1093/mnras/stab398},
	urldate = {2022-03-02},
	journal = {Monthly Notices of the Royal Astronomical Society},
	author = {Bonnerot, Clément and Lu, Wenbin and Hopkins, Philip F.},
	month = jul,
	year = {2021},
	note = {ADS Bibcode: 2021MNRAS.504.4885B},
	pages = {4885--4905},
}

@article{andalman_tidal_2022,
	title = {Tidal disruption discs formed and fed by stream-stream and stream-disc interactions in global {GRHD} simulations},
	volume = {510},
	issn = {0035-8711},
	url = {https://ui.adsabs.harvard.edu/abs/2022MNRAS.510.1627A},
	doi = {10.1093/mnras/stab3444},
	urldate = {2022-05-25},
	journal = {Monthly Notices of the Royal Astronomical Society},
	author = {Andalman, Zachary L. and Liska, Matthew T. P. and Tchekhovskoy, Alexander and Coughlin, Eric R. and Stone, Nicholas},
	month = feb,
	year = {2022},
	note = {ADS Bibcode: 2022MNRAS.510.1627A},
	pages = {1627--1648},
}

@article{piro_wind-reprocessed_2020,
	title = {Wind-reprocessed {Transients}},
	volume = {894},
	issn = {0004-637X},
	url = {https://ui.adsabs.harvard.edu/abs/2020ApJ...894....2P},
	doi = {10.3847/1538-4357/ab83f6},
	urldate = {2022-05-31},
	journal = {The Astrophysical Journal},
	author = {Piro, Anthony L. and Lu, Wenbin},
	month = may,
	year = {2020},
	note = {ADS Bibcode: 2020ApJ...894....2P},
	pages = {2},
}

@techreport{steinberg_origins_2022,
	title = {The {Origins} of {Peak} {Light} in {Tidal} {Disruption} {Events}},
	url = {https://ui.adsabs.harvard.edu/abs/2022arXiv220610641S},
	urldate = {2022-11-08},
	author = {Steinberg, Elad and Stone, Nicholas C.},
	month = jun,
	year = {2022},
	note = {Publication Title: arXiv e-prints
ADS Bibcode: 2022arXiv220610641S
Type: article},
}

@article{calderon_moving-mesh_2021,
	title = {Moving-mesh radiation-hydrodynamic simulations of wind-reprocessed transients},
	volume = {507},
	issn = {0035-8711, 1365-2966},
	url = {http://arxiv.org/abs/2105.08735},
	doi = {10.1093/mnras/stab2219},
	number = {1},
	urldate = {2022-12-22},
	journal = {Monthly Notices of the Royal Astronomical Society},
	author = {Calderón, Diego and Pejcha, Ondřej and Duffell, Paul C.},
	month = aug,
	year = {2021},
	note = {arXiv:2105.08735 [astro-ph]},
	pages = {1092--1105},
}

@article{jiang_infrared_2021,
	title = {Infrared {Echoes} of {Optical} {Tidal} {Disruption} {Events}: ∼1\% {Dust}-covering {Factor} or {Less} at {Subparsec} {Scale}},
	volume = {911},
	issn = {0004-637X},
	shorttitle = {Infrared {Echoes} of {Optical} {Tidal} {Disruption} {Events}},
	url = {https://ui.adsabs.harvard.edu/abs/2021ApJ...911...31J},
	doi = {10.3847/1538-4357/abe772},
	urldate = {2023-01-26},
	journal = {The Astrophysical Journal},
	author = {Jiang, Ning and Wang, Tinggui and Hu, Xueyang and Sun, Luming and Dou, Liming and Xiao, Lin},
	month = apr,
	year = {2021},
	note = {ADS Bibcode: 2021ApJ...911...31J},
	pages = {31},
}

@article{guillochon_hydrodynamical_2013,
	title = {Hydrodynamical {Simulations} to {Determine} the {Feeding} {Rate} of {Black} {Holes} by the {Tidal} {Disruption} of {Stars}: {The} {Importance} of the {Impact} {Parameter} and {Stellar} {Structure}},
	volume = {767},
	issn = {0004-637X},
	shorttitle = {Hydrodynamical {Simulations} to {Determine} the {Feeding} {Rate} of {Black} {Holes} by the {Tidal} {Disruption} of {Stars}},
	url = {https://ui.adsabs.harvard.edu/abs/2013ApJ...767...25G},
	doi = {10.1088/0004-637X/767/1/25},
	urldate = {2023-04-20},
	journal = {The Astrophysical Journal},
	author = {Guillochon, James and Ramirez-Ruiz, Enrico},
	month = apr,
	year = {2013},
	note = {ADS Bibcode: 2013ApJ...767...25G},
	pages = {25},
}

@techreport{huang_bright_2023,
	title = {A {Bright} {First} {Day} for {Tidal} {Disruption} {Event}},
	url = {https://ui.adsabs.harvard.edu/abs/2023arXiv230317443H},
	doi = {10.48550/arXiv.2303.17443},
	urldate = {2023-05-09},
	author = {Huang, Xiaoshan and Davis, Shane W. and Jiang, Yan-fei},
	month = mar,
	year = {2023},
	note = {Publication Title: arXiv e-prints
ADS Bibcode: 2023arXiv230317443H
Type: article},
}

@article{goodwin_radio-emitting_2023,
	title = {A radio-emitting outflow produced by the tidal disruption event {AT2020vwl}},
	issn = {0035-8711},
	url = {https://ui.adsabs.harvard.edu/abs/2023MNRAS.tmp.1199G},
	doi = {10.1093/mnras/stad1258},
	urldate = {2023-05-09},
	journal = {Monthly Notices of the Royal Astronomical Society},
	author = {Goodwin, A. J. and Alexander, K. D. and Miller-Jones, J. C. A. and Bietenholz, M. F. and van Velzen, S. and Anderson, G. E. and Berger, E. and Cendes, Y. and Chornock, R. and Coppejans, D. L. and Eftekhari, T. and Gezari, S. and Laskar, T. and Ramirez-Ruiz, E. and Saxton, R.},
	month = apr,
	year = {2023},
	note = {ADS Bibcode: 2023MNRAS.tmp.1199G},
}

@techreport{khatami_landscape_2023,
	title = {The {Landscape} of {Thermal} {Transients} from {Supernova} {Interacting} with a {Circumstellar} {Medium}},
	url = {https://ui.adsabs.harvard.edu/abs/2023arXiv230403360K},
	doi = {10.48550/arXiv.2304.03360},
	urldate = {2023-05-25},
	author = {Khatami, David and Kasen, Daniel},
	month = apr,
	year = {2023},
	note = {Publication Title: arXiv e-prints
ADS Bibcode: 2023arXiv230403360K
Type: article},
}

@article{hammerstein_final_2023,
	title = {The {Final} {Season} {Reimagined}: 30 {Tidal} {Disruption} {Events} from the {ZTF}-{I} {Survey}},
	volume = {942},
	issn = {0004-637X},
	shorttitle = {The {Final} {Season} {Reimagined}},
	url = {https://ui.adsabs.harvard.edu/abs/2023ApJ...942....9H},
	doi = {10.3847/1538-4357/aca283},
	urldate = {2023-06-07},
	journal = {The Astrophysical Journal},
	author = {Hammerstein, Erica and van Velzen, Sjoert and Gezari, Suvi and Cenko, S. Bradley and Yao, Yuhan and Ward, Charlotte and Frederick, Sara and Villanueva, Natalia and Somalwar, Jean J. and Graham, Matthew J. and Kulkarni, Shrinivas R. and Stern, Daniel and Andreoni, Igor and Bellm, Eric C. and Dekany, Richard and Dhawan, Suhail and Drake, Andrew J. and Fremling, Christoffer and Gatkine, Pradip and Groom, Steven L. and Ho, Anna Y. Q. and Kasliwal, Mansi M. and Karambelkar, Viraj and Kool, Erik C. and Masci, Frank J. and Medford, Michael S. and Perley, Daniel A. and Purdum, Josiah and van Roestel, Jan and Sharma, Yashvi and Sollerman, Jesper and Taggart, Kirsty and Yan, Lin},
	month = jan,
	year = {2023},
	note = {ADS Bibcode: 2023ApJ...942....9H},
	pages = {9},
}

@article{roth_x-ray_2016,
	title = {The {X}-{Ray} through {Optical} {Fluxes} and {Line} {Strengths} of {Tidal} {Disruption} {Events}},
	volume = {827},
	issn = {0004-637X},
	url = {https://ui.adsabs.harvard.edu/abs/2016ApJ...827....3R},
	doi = {10.3847/0004-637X/827/1/3},
	urldate = {2023-06-09},
	journal = {The Astrophysical Journal},
	author = {Roth, Nathaniel and Kasen, Daniel and Guillochon, James and Ramirez-Ruiz, Enrico},
	month = aug,
	year = {2016},
	note = {ADS Bibcode: 2016ApJ...827....3R},
	pages = {3},
}

@article{hammerstein_final_2023-1,
	title = {The {Final} {Season} {Reimagined}: 30 {Tidal} {Disruption} {Events} from the {ZTF}-{I} {Survey}},
	volume = {942},
	issn = {0004-637X},
	shorttitle = {The {Final} {Season} {Reimagined}},
	url = {https://ui.adsabs.harvard.edu/abs/2023ApJ...942....9H},
	doi = {10.3847/1538-4357/aca283},
	urldate = {2023-06-30},
	journal = {The Astrophysical Journal},
	author = {Hammerstein, Erica and van Velzen, Sjoert and Gezari, Suvi and Cenko, S. Bradley and Yao, Yuhan and Ward, Charlotte and Frederick, Sara and Villanueva, Natalia and Somalwar, Jean J. and Graham, Matthew J. and Kulkarni, Shrinivas R. and Stern, Daniel and Andreoni, Igor and Bellm, Eric C. and Dekany, Richard and Dhawan, Suhail and Drake, Andrew J. and Fremling, Christoffer and Gatkine, Pradip and Groom, Steven L. and Ho, Anna Y. Q. and Kasliwal, Mansi M. and Karambelkar, Viraj and Kool, Erik C. and Masci, Frank J. and Medford, Michael S. and Perley, Daniel A. and Purdum, Josiah and van Roestel, Jan and Sharma, Yashvi and Sollerman, Jesper and Taggart, Kirsty and Yan, Lin},
	month = jan,
	year = {2023},
	note = {ADS Bibcode: 2023ApJ...942....9H},
	pages = {9},
}

@misc{nicholl_at2022aedm_2023,
	title = {{AT2022aedm} and a new class of luminous, fast-cooling transients in elliptical galaxies},
	url = {https://ui.adsabs.harvard.edu/abs/2023arXiv230702556N},
	urldate = {2023-07-07},
	author = {Nicholl, M. and Srivastav, S. and Fulton, M. D. and Gomez, S. and Huber, M. E. and Oates, S. R. and Ramsden, P. and Rhodes, L. and Smartt, S. J. and Smith, K. W. and Aamer, A. and Anderson, J. P. and Bauer, F. E. and Berger, E. and de Boer, T. and Chambers, K. C. and Charalampopoulos, P. and Chen, T. -W. and Fender, R. P. and Fraser, M. and Gao, H. and Green, D. A. and Galbany, L. and Gompertz, B. P. and Gromadzki, M. and Gutiérrez, C. P. and Howell, D. A. and Inserra, C. and Jonker, P. G. and Kopsacheili, M. and Lowe, T. B. and Magnier, E. A. and McGee, S. L. and Moore, T. and Müller-Bravo, T. E. and Pessi, T. and Pursiainen, M. and Rest, A. and Ridley, E. J. and Shappee, B. J. and Sheng, X. and Smith, G. P. and Tucker, M. A. and Vinkó, J. and Wainscoat, R. J. and Wiseman, P. and Young, D. R.},
	month = jul,
	year = {2023},
	note = {Publication Title: arXiv e-prints
ADS Bibcode: 2023arXiv230702556N},
}

@article{thomsen_dynamical_2022-1,
	title = {Dynamical {Unification} of {Tidal} {Disruption} {Events}},
	volume = {937},
	issn = {0004-637X},
	url = {https://ui.adsabs.harvard.edu/abs/2022ApJ...937L..28T},
	doi = {10.3847/2041-8213/ac911f},
	urldate = {2023-07-14},
	journal = {The Astrophysical Journal},
	author = {Thomsen, Lars L. and Kwan, Tom M. and Dai, Lixin and Wu, Samantha C. and Roth, Nathaniel and Ramirez-Ruiz, Enrico},
	month = oct,
	year = {2022},
	note = {ADS Bibcode: 2022ApJ...937L..28T},
	pages = {L28},
}

@article{law-smith_stellar_2020,
	title = {Stellar {Tidal} {Disruption} {Events} with {Abundances} and {Realistic} {Structures} ({STARS}): {Library} of {Fallback} {Rates}},
	volume = {905},
	issn = {0004-637X},
	shorttitle = {Stellar {Tidal} {Disruption} {Events} with {Abundances} and {Realistic} {Structures} ({STARS})},
	url = {https://ui.adsabs.harvard.edu/abs/2020ApJ...905..141L},
	doi = {10.3847/1538-4357/abc489},
	urldate = {2023-08-07},
	journal = {The Astrophysical Journal},
	author = {Law-Smith, Jamie A. P. and Coulter, David A. and Guillochon, James and Mockler, Brenna and Ramirez-Ruiz, Enrico},
	month = dec,
	year = {2020},
	note = {ADS Bibcode: 2020ApJ...905..141L},
	pages = {141},
}

@article{kremer_wind-reprocessed_2023,
	title = {Wind-reprocessed transients from stellar-mass black hole {Tidal} {Disruption} {Events}},
	volume = {524},
	issn = {0035-8711},
	url = {https://ui.adsabs.harvard.edu/abs/2023MNRAS.524.6358K},
	doi = {10.1093/mnras/stad2239},
	urldate = {2023-08-16},
	journal = {Monthly Notices of the Royal Astronomical Society},
	author = {Kremer, Kyle and Mockler, Brenna and Piro, Anthony L. and Lombardi, James C.},
	month = oct,
	year = {2023},
	note = {ADS Bibcode: 2023MNRAS.524.6358K},
	pages = {6358--6373},
}

@misc{guolo_systematic_2023,
	title = {A systematic analysis of the {X}-ray emission in optically selected tidal disruption events: observational evidence for the unification of the optically and {X}-ray selected populations},
	shorttitle = {A systematic analysis of the {X}-ray emission in optically selected tidal disruption events},
	url = {https://ui.adsabs.harvard.edu/abs/2023arXiv230813019G},
	doi = {10.48550/arXiv.2308.13019},
	urldate = {2023-09-14},
	author = {Guolo, Muryel and Gezari, Suvi and Yao, Yuhan and van Velzen, Sjoert and Hammerstein, Erica and Cenko, S. Bradley and Tokayer, Yarone M.},
	month = aug,
	year = {2023},
	note = {Publication Title: arXiv e-prints
ADS Bibcode: 2023arXiv230813019G},
}

@article{mockler_energy_2021,
	title = {An {Energy} {Inventory} of {Tidal} {Disruption} {Events}},
	volume = {906},
	issn = {0004-637X},
	url = {https://ui.adsabs.harvard.edu/abs/2021ApJ...906..101M},
	doi = {10.3847/1538-4357/abc955},
	urldate = {2023-09-15},
	journal = {The Astrophysical Journal},
	author = {Mockler, Brenna and Ramirez-Ruiz, Enrico},
	month = jan,
	year = {2021},
	note = {ADS Bibcode: 2021ApJ...906..101M},
	pages = {101},
}

@article{mockler_weighing_2019,
	title = {Weighing {Black} {Holes} {Using} {Tidal} {Disruption} {Events}},
	volume = {872},
	issn = {0004-637X},
	url = {https://ui.adsabs.harvard.edu/abs/2019ApJ...872..151M},
	doi = {10.3847/1538-4357/ab010f},
	urldate = {2023-09-15},
	journal = {The Astrophysical Journal},
	author = {Mockler, Brenna and Guillochon, James and Ramirez-Ruiz, Enrico},
	month = feb,
	year = {2019},
	note = {ADS Bibcode: 2019ApJ...872..151M},
	pages = {151},
}

@article{ryu_tidal_2020,
	title = {Tidal {Disruptions} of {Main}-sequence {Stars}. {I}. {Observable} {Quantities} and {Their} {Dependence} on {Stellar} and {Black} {Hole} {Mass}},
	volume = {904},
	issn = {0004-637X},
	url = {https://ui.adsabs.harvard.edu/abs/2020ApJ...904...98R},
	doi = {10.3847/1538-4357/abb3cf},
	urldate = {2023-10-03},
	journal = {The Astrophysical Journal},
	author = {Ryu, Taeho and Krolik, Julian and Piran, Tsvi and Noble, Scott C.},
	month = dec,
	year = {2020},
	note = {ADS Bibcode: 2020ApJ...904...98R},
	pages = {98},
}

@article{jiang_prompt_2016,
	title = {Prompt {Radiation} and {Mass} {Outflows} from the {Stream}-{Stream} {Collisions} of {Tidal} {Disruption} {Events}},
	volume = {830},
	issn = {0004-637X},
	url = {https://ui.adsabs.harvard.edu/abs/2016ApJ...830..125J},
	doi = {10.3847/0004-637X/830/2/125},
	urldate = {2023-10-06},
	journal = {The Astrophysical Journal},
	author = {Jiang, Yan-Fei and Guillochon, James and Loeb, Abraham},
	month = oct,
	year = {2016},
	note = {ADS Bibcode: 2016ApJ...830..125J},
	pages = {125},
}

@article{parkinson_optical_2022,
	title = {Optical line spectra of tidal disruption events from reprocessing in optically thick outflows},
	volume = {510},
	issn = {0035-8711},
	url = {https://ui.adsabs.harvard.edu/abs/2022MNRAS.510.5426P},
	doi = {10.1093/mnras/stac027},
	urldate = {2023-10-10},
	journal = {Monthly Notices of the Royal Astronomical Society},
	author = {Parkinson, Edward J. and Knigge, Christian and Matthews, James H. and Long, Knox S. and Higginbottom, Nick and Sim, Stuart A. and Mangham, Samuel W.},
	month = mar,
	year = {2022},
	note = {ADS Bibcode: 2022MNRAS.510.5426P},
	pages = {5426--5443},
}

@misc{hammerstein_integral_2023,
	title = {Integral {Field} {Spectroscopy} of 13 {Tidal} {Disruption} {Event} {Hosts} from the {ZTF} {Survey}},
	url = {https://ui.adsabs.harvard.edu/abs/2023arXiv230715705H},
	doi = {10.48550/arXiv.2307.15705},
	urldate = {2023-10-16},
	author = {Hammerstein, Erica and Cenko, S. Bradley and Gezari, Suvi and Veilleux, Sylvain and O'Connor, Brendan and van Velzen, Sjoert and Ward, Charlotte and Yao, Yuhan and Graham, Matthew},
	month = jul,
	year = {2023},
	note = {Publication Title: arXiv e-prints
ADS Bibcode: 2023arXiv230715705H},
}

@article{blagorodnova_iptf16fnl_2017,
	title = {{iPTF16fnl}: {A} {Faint} and {Fast} {Tidal} {Disruption} {Event} in an {E}+{A} {Galaxy}},
	volume = {844},
	issn = {0004-637X},
	shorttitle = {{iPTF16fnl}},
	url = {https://ui.adsabs.harvard.edu/abs/2017ApJ...844...46B},
	doi = {10.3847/1538-4357/aa7579},
	urldate = {2023-12-14},
	journal = {The Astrophysical Journal},
	author = {Blagorodnova, N. and Gezari, S. and Hung, T. and Kulkarni, S. R. and Cenko, S. B. and Pasham, D. R. and Yan, L. and Arcavi, I. and Ben-Ami, S. and Bue, B. D. and Cantwell, T. and Cao, Y. and Castro-Tirado, A. J. and Fender, R. and Fremling, C. and Gal-Yam, A. and Ho, A. Y. Q. and Horesh, A. and Hosseinzadeh, G. and Kasliwal, M. M. and Kong, A. K. H. and Laher, R. R. and Leloudas, G. and Lunnan, R. and Masci, F. J. and Mooley, K. and Neill, J. D. and Nugent, P. and Powell, M. and Valeev, A. F. and Vreeswijk, P. M. and Walters, R. and Wozniak, P.},
	month = jul,
	year = {2017},
	note = {ADS Bibcode: 2017ApJ...844...46B},
	pages = {46},
}

@article{van_velzen_optical-ultraviolet_2020,
	title = {Optical-{Ultraviolet} {Tidal} {Disruption} {Events}},
	volume = {216},
	issn = {0038-6308},
	url = {https://ui.adsabs.harvard.edu/abs/2020SSRv..216..124V},
	doi = {10.1007/s11214-020-00753-z},
	urldate = {2023-12-15},
	journal = {Space Science Reviews},
	author = {van Velzen, Sjoert and Holoien, Thomas W. -S. and Onori, Francesca and Hung, Tiara and Arcavi, Iair},
	month = oct,
	year = {2020},
	note = {ADS Bibcode: 2020SSRv..216..124V},
	pages = {124},
}

@article{hinkle_discovery_2021,
	title = {Discovery and follow-up of {ASASSN}-19dj: an {X}-ray and {UV} luminous {TDE} in an extreme post-starburst galaxy},
	volume = {500},
	issn = {0035-8711},
	shorttitle = {Discovery and follow-up of {ASASSN}-19dj},
	url = {https://ui.adsabs.harvard.edu/abs/2021MNRAS.500.1673H},
	doi = {10.1093/mnras/staa3170},
	urldate = {2023-12-21},
	journal = {Monthly Notices of the Royal Astronomical Society},
	author = {Hinkle, Jason T. and Holoien, T. W. -S. and Auchettl, K. and Shappee, B. J. and Neustadt, J. M. M. and Payne, A. V. and Brown, J. S. and Kochanek, C. S. and Stanek, K. Z. and Graham, M. J. and Tucker, M. A. and Do, A. and Anderson, J. P. and Bose, S. and Chen, P. and Coulter, D. A. and Dimitriadis, G. and Dong, Subo and Foley, R. J. and Huber, M. E. and Hung, T. and Kilpatrick, C. D. and Pignata, G. and Piro, A. L. and Rojas-Bravo, C. and Siebert, M. R. and Stalder, B. and Thompson, Todd A. and Tonry, J. L. and Vallely, P. J. and Wisniewski, J. P.},
	month = jan,
	year = {2021},
	note = {ADS Bibcode: 2021MNRAS.500.1673H},
	pages = {1673--1696},
}

@article{charalampopoulos_detailed_2022,
	title = {A detailed spectroscopic study of tidal disruption events},
	volume = {659},
	issn = {0004-6361},
	url = {https://ui.adsabs.harvard.edu/abs/2022A&A...659A..34C},
	doi = {10.1051/0004-6361/202142122},
	urldate = {2024-02-05},
	journal = {Astronomy and Astrophysics},
	author = {Charalampopoulos, P. and Leloudas, G. and Malesani, D. B. and Wevers, T. and Arcavi, I. and Nicholl, M. and Pursiainen, M. and Lawrence, A. and Anderson, J. P. and Benetti, S. and Cannizzaro, G. and Chen, T. -W. and Galbany, L. and Gromadzki, M. and Gutiérrez, C. P. and Inserra, C. and Jonker, P. G. and Müller-Bravo, T. E. and Onori, F. and Short, P. and Sollerman, J. and Young, D. R.},
	month = mar,
	year = {2022},
	note = {ADS Bibcode: 2022A\&A...659A..34C},
	pages = {A34},
}

@article{short_tidal_2020,
	title = {The {Tidal} {Disruption} {Event} {AT} 2018hyz {I}: {Double}-peaked emission lines and a flat {Balmer} decrement},
	volume = {498},
	issn = {0035-8711, 1365-2966},
	shorttitle = {The {Tidal} {Disruption} {Event} {AT} 2018hyz {I}},
	url = {http://arxiv.org/abs/2003.05470},
	doi = {10.1093/mnras/staa2065},
	number = {3},
	urldate = {2024-02-05},
	journal = {Monthly Notices of the Royal Astronomical Society},
	author = {Short, P. and Nicholl, M. and Lawrence, A. and Gomez, S. and Arcavi, I. and Wevers, T. and Leloudas, G. and Schulze, S. and Anderson, J. P. and Berger, E. and Blanchard, P. K. and Burke, J. and Segura, N. Castro and Charalampopoulos, P. and Chornock, R. and Galbany, L. and Gromadzki, M. and Herzog, L. J. and Hiramatsu, D. and Horne, Keith and Hosseinzadeh, G. and Howell, D. Andrew and Ihanec, N. and Inserra, C. and Kankare, E. and Maguire, K. and McCully, C. and Bravo, T. E. Müller and Onori, F. and Sollerman, J. and Young, D. R.},
	month = sep,
	year = {2020},
	note = {arXiv:2003.05470 [astro-ph]},
	pages = {4119--4133},
}

@article{hung_double-peaked_2020,
	title = {Double-peaked {Balmer} {Emission} {Indicating} {Prompt} {Accretion} {Disk} {Formation} in an {X}-{Ray} {Faint} {Tidal} {Disruption} {Event}},
	volume = {903},
	issn = {0004-637X},
	url = {https://ui.adsabs.harvard.edu/abs/2020ApJ...903...31H},
	doi = {10.3847/1538-4357/abb606},
	urldate = {2024-02-05},
	journal = {The Astrophysical Journal},
	author = {Hung, Tiara and Foley, Ryan J. and Ramirez-Ruiz, Enrico and Dai, Jane L. and Auchettl, Katie and Kilpatrick, Charles D. and Mockler, Brenna and Brown, Jonathan S. and Coulter, David A. and Dimitriadis, Georgios and Holoien, Thomas W. -S. and Law-Smith, Jamie A. P. and Piro, Anthony L. and Rest, Armin and Rojas-Bravo, César and Siebert, Matthew R.},
	month = nov,
	year = {2020},
	note = {ADS Bibcode: 2020ApJ...903...31H},
	pages = {31},
}

@article{van_velzen_seventeen_2021,
	title = {Seventeen {Tidal} {Disruption} {Events} from the {First} {Half} of {ZTF} {Survey} {Observations}: {Entering} a {New} {Era} of {Population} {Studies}},
	volume = {908},
	issn = {0004-637X},
	shorttitle = {Seventeen {Tidal} {Disruption} {Events} from the {First} {Half} of {ZTF} {Survey} {Observations}},
	url = {https://ui.adsabs.harvard.edu/abs/2021ApJ...908....4V},
	doi = {10.3847/1538-4357/abc258},
	urldate = {2024-02-05},
	journal = {The Astrophysical Journal},
	author = {van Velzen, Sjoert and Gezari, Suvi and Hammerstein, Erica and Roth, Nathaniel and Frederick, Sara and Ward, Charlotte and Hung, Tiara and Cenko, S. Bradley and Stein, Robert and Perley, Daniel A. and Taggart, Kirsty and Foley, Ryan J. and Sollerman, Jesper and Blagorodnova, Nadejda and Andreoni, Igor and Bellm, Eric C. and Brinnel, Valery and De, Kishalay and Dekany, Richard and Feeney, Michael and Fremling, Christoffer and Giomi, Matteo and Golkhou, V. Zach and Graham, Matthew J. and Ho, Anna. Y. Q. and Kasliwal, Mansi M. and Kilpatrick, Charles D. and Kulkarni, Shrinivas R. and Kupfer, Thomas and Laher, Russ R. and Mahabal, Ashish and Masci, Frank J. and Miller, Adam A. and Nordin, Jakob and Riddle, Reed and Rusholme, Ben and van Santen, Jakob and Sharma, Yashvi and Shupe, David L. and Soumagnac, Maayane T.},
	month = feb,
	year = {2021},
	note = {ADS Bibcode: 2021ApJ...908....4V},
	pages = {4},
}

@article{masterson_new_2024,
	title = {A {New} {Population} of {Mid}-infrared-selected {Tidal} {Disruption} {Events}: {Implications} for {Tidal} {Disruption} {Event} {Rates} and {Host} {Galaxy} {Properties}},
	volume = {961},
	issn = {0004-637X},
	shorttitle = {A {New} {Population} of {Mid}-infrared-selected {Tidal} {Disruption} {Events}},
	url = {https://ui.adsabs.harvard.edu/abs/2024ApJ...961..211M},
	doi = {10.3847/1538-4357/ad18bb},
	urldate = {2024-02-07},
	journal = {The Astrophysical Journal},
	author = {Masterson, Megan and De, Kishalay and Panagiotou, Christos and Kara, Erin and Arcavi, Iair and Eilers, Anna-Christina and Frostig, Danielle and Gezari, Suvi and Grotova, Iuliia and Liu, Zhu and Malyali, Adam and Meisner, Aaron M. and Merloni, Andrea and Newsome, Megan and Rau, Arne and Simcoe, Robert A. and van Velzen, Sjoert},
	month = feb,
	year = {2024},
	note = {ADS Bibcode: 2024ApJ...961..211M},
	pages = {211},
}

@article{van_velzen_late-time_2019,
	title = {Late-time {UV} {Observations} of {Tidal} {Disruption} {Flares} {Reveal} {Unobscured}, {Compact} {Accretion} {Disks}},
	volume = {878},
	issn = {0004-637X},
	url = {https://ui.adsabs.harvard.edu/abs/2019ApJ...878...82V},
	doi = {10.3847/1538-4357/ab1844},
	urldate = {2024-02-23},
	journal = {The Astrophysical Journal},
	author = {van Velzen, Sjoert and Stone, Nicholas C. and Metzger, Brian D. and Gezari, Suvi and Brown, Thomas M. and Fruchter, Andrew S.},
	month = jun,
	year = {2019},
	note = {ADS Bibcode: 2019ApJ...878...82V},
	pages = {82},
}

@article{nicholl_outflow_2020,
	title = {An outflow powers the optical rise of the nearby, fast-evolving tidal disruption event {AT2019qiz}},
	volume = {499},
	issn = {0035-8711},
	url = {https://ui.adsabs.harvard.edu/abs/2020MNRAS.499..482N},
	doi = {10.1093/mnras/staa2824},
	urldate = {2024-02-27},
	journal = {Monthly Notices of the Royal Astronomical Society},
	author = {Nicholl, M. and Wevers, T. and Oates, S. R. and Alexander, K. D. and Leloudas, G. and Onori, F. and Jerkstrand, A. and Gomez, S. and Campana, S. and Arcavi, I. and Charalampopoulos, P. and Gromadzki, M. and Ihanec, N. and Jonker, P. G. and Lawrence, A. and Mandel, I. and Schulze, S. and Short, P. and Burke, J. and McCully, C. and Hiramatsu, D. and Howell, D. A. and Pellegrino, C. and Abbot, H. and Anderson, J. P. and Berger, E. and Blanchard, P. K. and Cannizzaro, G. and Chen, T. -W. and Dennefeld, M. and Galbany, L. and González-Gaitán, S. and Hosseinzadeh, G. and Inserra, C. and Irani, I. and Kuin, P. and Müller-Bravo, T. and Pineda, J. and Ross, N. P. and Roy, R. and Smartt, S. J. and Smith, K. W. and Tucker, B. and Wyrzykowski, Ł. and Young, D. R.},
	month = nov,
	year = {2020},
	note = {ADS Bibcode: 2020MNRAS.499..482N},
	pages = {482--504},
}

@article{hu_optical_2024,
	title = {Optical {Appearance} of {Eccentric} {Tidal} {Disruption} {Events}},
	volume = {963},
	issn = {0004-637X},
	url = {https://ui.adsabs.harvard.edu/abs/2024ApJ...963L..27H},
	doi = {10.3847/2041-8213/ad29ec},
	urldate = {2024-03-06},
	journal = {The Astrophysical Journal},
	author = {Hu, Fangyi (Fitz) and Price, Daniel J. and Mandel, Ilya},
	month = mar,
	year = {2024},
	note = {ADS Bibcode: 2024ApJ...963L..27H},
	pages = {L27},
}

@misc{wang_identifying_2024,
	title = {Identifying changing-look {AGNs} using variability characteristics},
	url = {https://ui.adsabs.harvard.edu/abs/2024arXiv240218131W},
	doi = {10.48550/arXiv.2402.18131},
	urldate = {2024-03-15},
	author = {Wang, Shu and Woo, Jong-Hak and Gallo, Elena and Guo, Hengxiao and Son, Donghoon and Kong, Minzhi and Mandal, Amit Kumar and Cho, Hojin and Kim, Changseok and Shin, Jaejin},
	month = feb,
	year = {2024},
	note = {Publication Title: arXiv e-prints
ADS Bibcode: 2024arXiv240218131W},
}

@article{ryu_shocks_2023,
	title = {Shocks {Power} {Tidal} {Disruption} {Events}},
	volume = {957},
	issn = {0004-637X},
	url = {https://ui.adsabs.harvard.edu/abs/2023ApJ...957...12R},
	doi = {10.3847/1538-4357/acf5de},
	urldate = {2024-04-08},
	journal = {The Astrophysical Journal},
	author = {Ryu, Taeho and Krolik, Julian and Piran, Tsvi and Noble, Scott C. and Avara, Mark},
	month = nov,
	year = {2023},
	note = {ADS Bibcode: 2023ApJ...957...12R},
	pages = {12},
}

@article{lu_missing_2018,
	title = {On the {Missing} {Energy} {Puzzle} of {Tidal} {Disruption} {Events}},
	volume = {865},
	issn = {0004-637X},
	url = {https://ui.adsabs.harvard.edu/abs/2018ApJ...865..128L},
	doi = {10.3847/1538-4357/aad54a},
	urldate = {2024-04-10},
	journal = {The Astrophysical Journal},
	author = {Lu, Wenbin and Kumar, Pawan},
	month = oct,
	year = {2018},
	note = {ADS Bibcode: 2018ApJ...865..128L},
	pages = {128},
}

@article{wevers_elliptical_2022,
	title = {An elliptical accretion disk following the tidal disruption event {AT} 2020zso},
	volume = {666},
	issn = {0004-6361},
	url = {https://ui.adsabs.harvard.edu/abs/2022A&A...666A...6W},
	doi = {10.1051/0004-6361/202142616},
	urldate = {2024-04-16},
	journal = {Astronomy and Astrophysics},
	author = {Wevers, T. and Nicholl, M. and Guolo, M. and Charalampopoulos, P. and Gromadzki, M. and Reynolds, T. M. and Kankare, E. and Leloudas, G. and Anderson, J. P. and Arcavi, I. and Cannizzaro, G. and Chen, T. -W. and Ihanec, N. and Inserra, C. and Gutiérrez, C. P. and Jonker, P. G. and Lawrence, A. and Magee, M. R. and Müller-Bravo, T. E. and Onori, F. and Ridley, E. and Schulze, S. and Short, P. and Hiramatsu, D. and Newsome, M. and Terwel, J. H. and Yang, S. and Young, D.},
	month = oct,
	year = {2022},
	note = {ADS Bibcode: 2022A\&A...666A...6W},
	pages = {A6},
}

@misc{price_eddington_2024,
	title = {Eddington envelopes: {The} fate of stars on parabolic orbits tidally disrupted by supermassive black holes},
	shorttitle = {Eddington envelopes},
	url = {https://ui.adsabs.harvard.edu/abs/2024arXiv240409381P},
	doi = {10.48550/arXiv.2404.09381},
	urldate = {2024-04-24},
	author = {Price, Daniel J. and Liptai, David and Mandel, Ilya and Shepherd, Joanna and Lodato, Giuseppe and Levin, Yuri},
	month = apr,
	year = {2024},
	note = {Publication Title: arXiv e-prints
ADS Bibcode: 2024arXiv240409381P},
}

@article{hung_revisiting_2017,
	title = {Revisiting {Optical} {Tidal} {Disruption} {Events} with {iPTF16axa}},
	volume = {842},
	issn = {0004-637X},
	url = {https://ui.adsabs.harvard.edu/abs/2017ApJ...842...29H},
	doi = {10.3847/1538-4357/aa7337},
	urldate = {2024-05-14},
	journal = {The Astrophysical Journal},
	publisher = {IOP},
	author = {Hung, T. and Gezari, S. and Blagorodnova, N. and Roth, N. and Cenko, S. B. and Kulkarni, S. R. and Horesh, A. and Arcavi, I. and McCully, C. and Yan, Lin and Lunnan, R. and Fremling, C. and Cao, Y. and Nugent, P. E. and Wozniak, P.},
	month = jun,
	year = {2017},
	note = {ADS Bibcode: 2017ApJ...842...29H},
	pages = {29},
}

@article{dai_unified_2018,
	title = {A {Unified} {Model} for {Tidal} {Disruption} {Events}},
	volume = {859},
	issn = {0004-637X},
	url = {https://ui.adsabs.harvard.edu/abs/2018ApJ...859L..20D},
	doi = {10.3847/2041-8213/aab429},
	urldate = {2024-06-14},
	journal = {The Astrophysical Journal},
	publisher = {IOP},
	author = {Dai, Lixin and McKinney, Jonathan C. and Roth, Nathaniel and Ramirez-Ruiz, Enrico and Miller, M. Coleman},
	month = jun,
	year = {2018},
	note = {ADS Bibcode: 2018ApJ...859L..20D},
	pages = {L20},
}

@article{roth_what_2018,
	title = {What {Sets} the {Line} {Profiles} in {Tidal} {Disruption} {Events}?},
	volume = {855},
	issn = {0004-637X},
	url = {https://ui.adsabs.harvard.edu/abs/2018ApJ...855...54R},
	doi = {10.3847/1538-4357/aaaec6},
	urldate = {2024-06-14},
	journal = {The Astrophysical Journal},
	publisher = {IOP},
	author = {Roth, Nathaniel and Kasen, Daniel},
	month = mar,
	year = {2018},
	note = {ADS Bibcode: 2018ApJ...855...54R},
	pages = {54},
}

@article{kasen_time-dependent_2006,
	title = {Time-dependent {Monte} {Carlo} {Radiative} {Transfer} {Calculations} for {Three}-dimensional {Supernova} {Spectra}, {Light} {Curves}, and {Polarization}},
	volume = {651},
	issn = {0004-637X},
	url = {https://ui.adsabs.harvard.edu/abs/2006ApJ...651..366K},
	doi = {10.1086/506190},
	urldate = {2024-07-11},
	journal = {The Astrophysical Journal},
	publisher = {IOP},
	author = {Kasen, Daniel and Thomas, R. C. and Nugent, P.},
	month = nov,
	year = {2006},
	note = {ADS Bibcode: 2006ApJ...651..366K},
	pages = {366--380},
}

@misc{cendes_ubiquitous_2023,
	title = {Ubiquitous {Late} {Radio} {Emission} from {Tidal} {Disruption} {Events}},
	url = {https://ui.adsabs.harvard.edu/abs/2023arXiv230813595C},
	doi = {10.48550/arXiv.2308.13595},
	urldate = {2024-07-24},
	author = {Cendes, Yvette and Berger, Edo and Alexander, Kate D. and Chornock, Ryan and Margutti, Raffaella and Metzger, Brian and Wieringa, Mark H. and Bietenholz, Michael F. and Hajela, Aprajita and Laskar, Tanmoy and Stroh, Michael C. and Terreran, Giacomo},
	month = aug,
	year = {2023},
	note = {Publication Title: arXiv e-prints
ADS Bibcode: 2023arXiv230813595C},
}

@article{van_velzen_reverberation_2021,
	title = {Reverberation in {Tidal} {Disruption} {Events}: {Dust} {Echoes}, {Coronal} {Emission} {Lines}, {Multi}-wavelength {Cross}-correlations, and {QPOs}},
	volume = {217},
	issn = {0038-6308},
	shorttitle = {Reverberation in {Tidal} {Disruption} {Events}},
	url = {https://ui.adsabs.harvard.edu/abs/2021SSRv..217...63V},
	doi = {10.1007/s11214-021-00835-6},
	urldate = {2024-10-12},
	journal = {Space Science Reviews},
	author = {van Velzen, Sjoert and Pasham, Dheeraj R. and Komossa, Stefanie and Yan, Lin and Kara, Erin A.},
	month = aug,
	year = {2021},
	note = {ADS Bibcode: 2021SSRv..217...63V},
	pages = {63},
}

@article{matthee_little_2024,
	title = {Little {Red} {Dots}: {An} {Abundant} {Population} of {Faint} {Active} {Galactic} {Nuclei} at z ∼ 5 {Revealed} by the {EIGER} and {FRESCO} {JWST} {Surveys}},
	volume = {963},
	issn = {0004-637X},
	shorttitle = {Little {Red} {Dots}},
	url = {https://ui.adsabs.harvard.edu/abs/2024ApJ...963..129M},
	doi = {10.3847/1538-4357/ad2345},
	urldate = {2024-10-12},
	journal = {The Astrophysical Journal},
	publisher = {IOP},
	author = {Matthee, Jorryt and Naidu, Rohan P. and Brammer, Gabriel and Chisholm, John and Eilers, Anna-Christina and Goulding, Andy and Greene, Jenny and Kashino, Daichi and Labbe, Ivo and Lilly, Simon J. and Mackenzie, Ruari and Oesch, Pascal A. and Weibel, Andrea and Wuyts, Stijn and Xiao, Mengyuan and Bordoloi, Rongmon and Bouwens, Rychard and van Dokkum, Pieter and Illingworth, Garth and Kramarenko, Ivan and Maseda, Michael V. and Mason, Charlotte and Meyer, Romain A. and Nelson, Erica J. and Reddy, Naveen A. and Shivaei, Irene and Simcoe, Robert A. and Yue, Minghao},
	month = mar,
	year = {2024},
	note = {ADS Bibcode: 2024ApJ...963..129M},
	pages = {129},
}

@article{mockler_tidal_2024,
	title = {Tidal {Disruption} {Events} from {Stripped} {Stars}},
	volume = {973},
	issn = {0004-637X},
	url = {https://ui.adsabs.harvard.edu/abs/2024ApJ...973L...9M},
	doi = {10.3847/2041-8213/ad6c34},
	urldate = {2024-10-15},
	journal = {The Astrophysical Journal},
	publisher = {IOP},
	author = {Mockler, Brenna and Gallegos-Garcia, Monica and Götberg, Ylva and Miller, Jon M. and Ramirez-Ruiz, Enrico},
	month = sep,
	year = {2024},
	note = {ADS Bibcode: 2024ApJ...973L...9M},
	pages = {L9},
}

@article{steinberg_stream-disk_2024,
	title = {Stream-disk shocks as the origins of peak light in tidal disruption events},
	volume = {625},
	issn = {0028-0836},
	url = {https://ui.adsabs.harvard.edu/abs/2024Natur.625..463S},
	doi = {10.1038/s41586-023-06875-y},
	urldate = {2024-10-15},
	journal = {Nature},
	author = {Steinberg, Elad and Stone, Nicholas C.},
	month = jan,
	year = {2024},
	note = {ADS Bibcode: 2024Natur.625..463S},
	pages = {463--467},
}

@article{price_eddington_2024-1,
	title = {Eddington {Envelopes}: {The} {Fate} of {Stars} on {Parabolic} {Orbits} {Tidally} {Disrupted} by {Supermassive} {Black} {Holes}},
	volume = {971},
	issn = {0004-637X},
	shorttitle = {Eddington {Envelopes}},
	url = {https://ui.adsabs.harvard.edu/abs/2024ApJ...971L..46P},
	doi = {10.3847/2041-8213/ad6862},
	urldate = {2024-10-15},
	journal = {The Astrophysical Journal},
	publisher = {IOP},
	author = {Price, Daniel J. and Liptai, David and Mandel, Ilya and Shepherd, Joanna and Lodato, Giuseppe and Levin, Yuri},
	month = aug,
	year = {2024},
	note = {ADS Bibcode: 2024ApJ...971L..46P},
	pages = {L46},
}

@article{malyali_transient_2024,
	title = {Transient fading {X}-ray emission detected during the optical rise of a tidal disruption event},
	volume = {531},
	issn = {0035-8711},
	url = {https://ui.adsabs.harvard.edu/abs/2024MNRAS.531.1256M},
	doi = {10.1093/mnras/stae927},
	urldate = {2024-10-27},
	journal = {Monthly Notices of the Royal Astronomical Society},
	publisher = {OUP},
	author = {Malyali, A. and Rau, A. and Bonnerot, C. and Goodwin, A. J. and Liu, Z. and Anderson, G. E. and Brink, J. and Buckley, D. A. H. and Merloni, A. and Miller-Jones, J. C. A. and Grotova, I. and Kawka, A.},
	month = jun,
	year = {2024},
	note = {ADS Bibcode: 2024MNRAS.531.1256M},
	pages = {1256--1275},
}

@article{huang_pre-peak_2024,
	title = {Pre-peak {Emission} in {Tidal} {Disruption} {Events}},
	volume = {974},
	issn = {0004-637X},
	url = {https://ui.adsabs.harvard.edu/abs/2024ApJ...974..165H},
	doi = {10.3847/1538-4357/ad6c39},
	urldate = {2024-10-29},
	journal = {The Astrophysical Journal},
	publisher = {IOP},
	author = {Huang, Xiaoshan and Davis, Shane W. and Jiang, Yan-fei},
	month = oct,
	year = {2024},
	note = {ADS Bibcode: 2024ApJ...974..165H},
	pages = {165},
}

@article{jankovic_spin-induced_2024,
	title = {Spin-induced offset stream self-crossing shocks in tidal disruption events},
	volume = {529},
	issn = {0035-8711},
	url = {https://ui.adsabs.harvard.edu/abs/2024MNRAS.529..673J},
	doi = {10.1093/mnras/stae580},
	urldate = {2025-03-06},
	journal = {Monthly Notices of the Royal Astronomical Society},
	publisher = {OUP},
	author = {Jankovič, T. and Bonnerot, C. and Gomboc, A.},
	month = mar,
	year = {2024},
	note = {ADS Bibcode: 2024MNRAS.529..673J},
	pages = {673--687},
}

@article{yao_tidal_2023-1,
	title = {Tidal {Disruption} {Event} {Demographics} with the {Zwicky} {Transient} {Facility}: {Volumetric} {Rates}, {Luminosity} {Function}, and {Implications} for the {Local} {Black} {Hole} {Mass} {Function}},
	volume = {955},
	issn = {0004-637X},
	shorttitle = {Tidal {Disruption} {Event} {Demographics} with the {Zwicky} {Transient} {Facility}},
	url = {https://ui.adsabs.harvard.edu/abs/2023ApJ...955L...6Y},
	doi = {10.3847/2041-8213/acf216},
	urldate = {2025-04-08},
	journal = {The Astrophysical Journal},
	publisher = {IOP},
	author = {Yao, Yuhan and Ravi, Vikram and Gezari, Suvi and van Velzen, Sjoert and Lu, Wenbin and Schulze, Steve and Somalwar, Jean J. and Kulkarni, S. R. and Hammerstein, Erica and Nicholl, Matt and Graham, Matthew J. and Perley, Daniel A. and Cenko, S. Bradley and Stein, Robert and Ricarte, Angelo and Chadayammuri, Urmila and Quataert, Eliot and Bellm, Eric C. and Bloom, Joshua S. and Dekany, Richard and Drake, Andrew J. and Groom, Steven L. and Mahabal, Ashish A. and Prince, Thomas A. and Riddle, Reed and Rusholme, Ben and Sharma, Yashvi and Sollerman, Jesper and Yan, Lin},
	month = sep,
	year = {2023},
	note = {ADS Bibcode: 2023ApJ...955L...6Y},
	pages = {L6},
}

@article{van_velzen_discovery_2016,
	title = {Discovery of {Transient} {Infrared} {Emission} from {Dust} {Heated} by {Stellar} {Tidal} {Disruption} {Flares}},
	volume = {829},
	issn = {0004-637X},
	url = {https://ui.adsabs.harvard.edu/abs/2016ApJ...829...19V},
	doi = {10.3847/0004-637X/829/1/19},
	urldate = {2025-05-02},
	journal = {The Astrophysical Journal},
	publisher = {IOP},
	author = {van Velzen, S. and Mendez, A. J. and Krolik, J. H. and Gorjian, V.},
	month = sep,
	year = {2016},
	note = {ADS Bibcode: 2016ApJ...829...19V},
	pages = {19},
}

@article{leloudas_spectral_2019,
	title = {The {Spectral} {Evolution} of {AT} 2018dyb and the {Presence} of {Metal} {Lines} in {Tidal} {Disruption} {Events}},
	volume = {887},
	issn = {0004-637X},
	url = {https://ui.adsabs.harvard.edu/abs/2019ApJ...887..218L},
	doi = {10.3847/1538-4357/ab5792},
	urldate = {2025-06-17},
	journal = {The Astrophysical Journal},
	publisher = {IOP},
	author = {Leloudas, Giorgos and Dai, Lixin and Arcavi, Iair and Vreeswijk, Paul M. and Mockler, Brenna and Roy, Rupak and Malesani, Daniele B. and Schulze, Steve and Wevers, Thomas and Fraser, Morgan and Ramirez-Ruiz, Enrico and Auchettl, Katie and Burke, Jamison and Cannizzaro, Giacomo and Charalampopoulos, Panos and Chen, Ting-Wan and Cikota, Aleksandar and Della Valle, Massimo and Galbany, Lluis and Gromadzki, Mariusz and Heintz, Kasper E. and Hiramatsu, Daichi and Jonker, Peter G. and Kostrzewa-Rutkowska, Zuzanna and Maguire, Kate and Mandel, Ilya and Nicholl, Matt and Onori, Francesca and Roth, Nathaniel and Smartt, Stephen J. and Wyrzykowski, Lukasz and Young, Dave R.},
	month = dec,
	year = {2019},
	note = {ADS Bibcode: 2019ApJ...887..218L},
	pages = {218},
}

@article{goodwin_at2019azh_2022,
	title = {{AT2019azh}: an unusually long-lived, radio-bright thermal tidal disruption event},
	volume = {511},
	issn = {0035-8711},
	shorttitle = {{AT2019azh}},
	url = {https://ui.adsabs.harvard.edu/abs/2022MNRAS.511.5328G},
	doi = {10.1093/mnras/stac333},
	urldate = {2025-06-23},
	journal = {Monthly Notices of the Royal Astronomical Society},
	publisher = {OUP},
	author = {Goodwin, A. J. and van Velzen, S. and Miller-Jones, J. C. A. and Mummery, A. and Bietenholz, M. F. and Wederfoort, A. and Hammerstein, E. and Bonnerot, C. and Hoffmann, J. and Yan, L.},
	month = apr,
	year = {2022},
	note = {ADS Bibcode: 2022MNRAS.511.5328G},
	pages = {5328--5345},
}

@article{cao_tidal_2024,
	title = {Tidal {Disruption} {Event} {AT2020ocn}: {Early} {Time} {X}-{Ray} {Flares} {Caused} by a {Possible} {Disk} {Alignment} {Process}},
	volume = {970},
	issn = {0004-637X},
	shorttitle = {Tidal {Disruption} {Event} {AT2020ocn}},
	url = {https://ui.adsabs.harvard.edu/abs/2024ApJ...970...89C},
	doi = {10.3847/1538-4357/ad496f},
	urldate = {2025-06-23},
	journal = {The Astrophysical Journal},
	publisher = {IOP},
	author = {Cao, Z. and Jonker, P. G. and Pasham, D. R. and Wen, S. and Stone, N. C. and Zabludoff, A. I.},
	month = jul,
	year = {2024},
	note = {ADS Bibcode: 2024ApJ...970...89C},
	pages = {89},
}

@misc{zhang_radiation_2025,
	title = {Radiation {GRMHD} {Models} of {Accretion} onto {Stellar}-{Mass} {Black} {Holes}: {I}. {Survey} of {Eddington} {Ratios}},
	shorttitle = {Radiation {GRMHD} {Models} of {Accretion} onto {Stellar}-{Mass} {Black} {Holes}},
	url = {https://ui.adsabs.harvard.edu/abs/2025arXiv250602289Z},
	doi = {10.48550/arXiv.2506.02289},
	urldate = {2025-07-14},
	publisher = {arXiv},
	author = {Zhang, Lizhong and Stone, James M. and Mullen, Patrick D. and Davis, Shane W. and Jiang, Yan-Fei and White, Christopher J.},
	month = jun,
	year = {2025},
	note = {ADS Bibcode: 2025arXiv250602289Z},
}

@article{khatami_landscape_2024,
	title = {The {Landscape} of {Thermal} {Transients} from {Supernovae} {Interacting} with a {Circumstellar} {Medium}},
	volume = {972},
	issn = {0004-637X},
	url = {https://ui.adsabs.harvard.edu/abs/2024ApJ...972..140K},
	doi = {10.3847/1538-4357/ad60c0},
	urldate = {2025-09-03},
	journal = {The Astrophysical Journal},
	author = {Khatami, David K. and Kasen, Daniel N.},
	month = sep,
	year = {2024},
	note = {ADS Bibcode: 2024ApJ...972..140K},
	pages = {140},
}

@misc{alexander_multi-wavelength_2025,
	title = {The {Multi}-{Wavelength} {Context} of {Delayed} {Radio} {Emission} in {TDEs}: {Evidence} for {Accretion}-{Driven} {Outflows}},
	shorttitle = {The {Multi}-{Wavelength} {Context} of {Delayed} {Radio} {Emission} in {TDEs}},
	url = {https://ui.adsabs.harvard.edu/abs/2025arXiv250612729A},
	doi = {10.48550/arXiv.2506.12729},
	urldate = {2025-10-01},
	publisher = {arXiv},
	author = {Alexander, Kate D. and Margutti, Raffaella and Gomez, Sebastian and Stroh, Michael and Chornock, Ryan and Laskar, Tanmoy and Cendes, Y. and Berger, Edo and Eftekhari, Tarraneh and Franz, Noah and Hajela, Aprajita and Metzger, B. D. and Terreran, Giacomo and Bietenholz, Michael and Christy, Collin and de Colle, Fabio and Komossa, S. and Nicholl, Matt and Ramirez-Ruiz, Enrico and Saxton, Richard and Schroeder, Genevieve and Williams, Peter and Wu, William},
	month = jun,
	year = {2025},
	note = {ADS Bibcode: 2025arXiv250612729A},
}

@article{cao_tidal_2024-1,
	title = {Tidal {Disruption} {Event} {AT2020ocn}: {Early} {Time} {X}-{Ray} {Flares} {Caused} by a {Possible} {Disk} {Alignment} {Process}},
	volume = {970},
	issn = {0004-637X},
	shorttitle = {Tidal {Disruption} {Event} {AT2020ocn}},
	url = {https://ui.adsabs.harvard.edu/abs/2024ApJ...970...89C},
	doi = {10.3847/1538-4357/ad496f},
	urldate = {2025-11-21},
	journal = {The Astrophysical Journal},
	publisher = {IOP},
	author = {Cao, Z. and Jonker, P. G. and Pasham, D. R. and Wen, S. and Stone, N. C. and Zabludoff, A. I.},
	month = jul,
	year = {2024},
	note = {ADS Bibcode: 2024ApJ...970...89C},
	pages = {89},
}

@article{schlafly_measuring_2011,
	title = {Measuring {Reddening} with {Sloan} {Digital} {Sky} {Survey} {Stellar} {Spectra} and {Recalibrating} {SFD}},
	volume = {737},
	issn = {0004-637X},
	url = {https://ui.adsabs.harvard.edu/abs/2011ApJ...737..103S},
	doi = {10.1088/0004-637X/737/2/103},
	urldate = {2025-11-21},
	journal = {The Astrophysical Journal},
	publisher = {IOP},
	author = {Schlafly, Edward F. and Finkbeiner, Douglas P.},
	month = aug,
	year = {2011},
	note = {ADS Bibcode: 2011ApJ...737..103S},
	pages = {103},
}

@misc{andalman_resolving_2025,
	title = {Resolving the ({Debate} {About}) {Nozzle} {Shocks} in {Tidal} {Disruption} {Events}},
	url = {https://ui.adsabs.harvard.edu/abs/2025arXiv251208928A},
	doi = {10.48550/arXiv.2512.08928},
	urldate = {2025-12-12},
	publisher = {arXiv},
	author = {Andalman, Zachary L. and Quataert, Eliot and Coughlin, Eric R. and Nixon, C. J.},
	month = dec,
	year = {2025},
	note = {ADS Bibcode: 2025arXiv251208928A},
}

@book{mihalas_stellar_1978,
	title = {Stellar atmospheres},
	url = {https://ui.adsabs.harvard.edu/abs/1978stat.book.....M},
	urldate = {2025-12-18},
	author = {Mihalas, Dimitri},
	month = jan,
	year = {1978},
	note = {Publication Title: San Francisco: W.H. Freeman
ADS Bibcode: 1978stat.book.....M},
}

@misc{huang_x-ray_2025,
	title = {X-ray {Variability} and {Photosphere} {Evolution} during {Accretion} {Disk} {Formation} in {Tidal} {Disruption} {Events}},
	url = {https://ui.adsabs.harvard.edu/abs/2025arXiv251212985H},
	doi = {10.48550/arXiv.2512.12985},
	urldate = {2025-12-18},
	publisher = {arXiv},
	author = {Huang, Xiaoshan and Meza, Maria Renee and Yun, Sol Bin and Mockler, Brenna and Davis, Shane W. and Jiang, Yan-fei},
	month = dec,
	year = {2025},
	note = {ADS Bibcode: 2025arXiv251212985H},
}

@article{hinkle_mid-infrared_2024,
	title = {Mid-infrared echoes of ambiguous nuclear transients reveal high dust covering fractions: evidence for dusty tori},
	volume = {531},
	issn = {0035-8711},
	shorttitle = {Mid-infrared echoes of ambiguous nuclear transients reveal high dust covering fractions},
	url = {https://ui.adsabs.harvard.edu/abs/2024MNRAS.531.2603H},
	doi = {10.1093/mnras/stae1229},
	urldate = {2025-12-18},
	journal = {Monthly Notices of the Royal Astronomical Society},
	publisher = {OUP},
	author = {Hinkle, Jason T.},
	month = jun,
	year = {2024},
	note = {ADS Bibcode: 2024MNRAS.531.2603H},
	pages = {2603--2614},
}

@article{roth_x-ray_2016-1,
	title = {The {X}-{Ray} through {Optical} {Fluxes} and {Line} {Strengths} of {Tidal} {Disruption} {Events}},
	volume = {827},
	issn = {0004-637X},
	url = {https://ui.adsabs.harvard.edu/abs/2016ApJ...827....3R},
	doi = {10.3847/0004-637X/827/1/3},
	urldate = {2025-12-18},
	journal = {The Astrophysical Journal},
	publisher = {IOP},
	author = {Roth, Nathaniel and Kasen, Daniel and Guillochon, James and Ramirez-Ruiz, Enrico},
	month = aug,
	year = {2016},
	note = {ADS Bibcode: 2016ApJ...827....3R},
	pages = {3},
}

@article{jones_intrinsic_2016,
	title = {The {Intrinsic} {Eddington} {Ratio} {Distribution} of {Active} {Galactic} {Nuclei} in {Star}-forming {Galaxies} from the {Sloan} {Digital} {Sky} {Survey}},
	volume = {826},
	issn = {0004-637X},
	url = {https://ui.adsabs.harvard.edu/abs/2016ApJ...826...12J},
	doi = {10.3847/0004-637X/826/1/12},
	urldate = {2025-12-23},
	journal = {The Astrophysical Journal},
	publisher = {IOP},
	author = {Jones, Mackenzie L. and Hickox, Ryan C. and Black, Christine S. and Hainline, Kevin N. and DiPompeo, Michael A. and Goulding, Andy D.},
	month = jul,
	year = {2016},
	note = {ADS Bibcode: 2016ApJ...826...12J},
	pages = {12},
}

@misc{blanton_eddington_2025,
	title = {The {Eddington} {Ratio} {Distribution} of {Narrow} {Line} {Active} {Galactic} {Nuclei}},
	url = {https://ui.adsabs.harvard.edu/abs/2025arXiv251114575B},
	doi = {10.48550/arXiv.2511.14575},
	urldate = {2025-12-23},
	publisher = {arXiv},
	author = {Blanton, Michael R. and Suresh, Arjun and Westfall, Kyle B. and Liu, Dou and Moustakas, John},
	month = nov,
	year = {2025},
	note = {ADS Bibcode: 2025arXiv251114575B},
}

@article{jiang_super-eddington_2019,
	title = {Super-{Eddington} {Accretion} {Disks} around {Supermassive} {Black} {Holes}},
	volume = {880},
	issn = {0004-637X},
	url = {https://ui.adsabs.harvard.edu/abs/2019ApJ...880...67J},
	doi = {10.3847/1538-4357/ab29ff},
	urldate = {2026-01-13},
	journal = {The Astrophysical Journal},
	publisher = {IOP},
	author = {Jiang, Yan-Fei and Stone, James M. and Davis, Shane W.},
	month = aug,
	year = {2019},
	note = {ADS Bibcode: 2019ApJ...880...67J},
	pages = {67},
}

@article{dodd_mid-infrared_2023,
	title = {Mid-infrared {Outbursts} in {Nearby} {Galaxies}: {Nuclear} {Obscuration} and {Connections} to {Hidden} {Tidal} {Disruption} {Events} and {Changing}-look {Active} {Galactic} {Nuclei}},
	volume = {959},
	issn = {0004-637X},
	shorttitle = {Mid-infrared {Outbursts} in {Nearby} {Galaxies}},
	url = {https://ui.adsabs.harvard.edu/abs/2023ApJ...959L..19D},
	doi = {10.3847/2041-8213/ad1112},
	urldate = {2026-01-13},
	journal = {The Astrophysical Journal},
	publisher = {IOP},
	author = {Dodd, Sierra A. and Nukala, Arya and Connor, Isabelle and Auchettl, Katie and French, K. D. and Law-Smith, Jamie A. P. and Hammerstein, Erica and Ramirez-Ruiz, Enrico},
	month = dec,
	year = {2023},
	note = {ADS Bibcode: 2023ApJ...959L..19D},
	pages = {L19},
}

@article{hinkle_examining_2020,
	title = {Examining a {Peak}-luminosity/{Decline}-rate {Relationship} for {Tidal} {Disruption} {Events}},
	volume = {894},
	issn = {0004-637X},
	url = {https://ui.adsabs.harvard.edu/abs/2020ApJ...894L..10H},
	doi = {10.3847/2041-8213/ab89a2},
	urldate = {2026-01-13},
	journal = {The Astrophysical Journal},
	publisher = {IOP},
	author = {Hinkle, Jason T. and Holoien, Thomas W.-S. and Shappee, Benjamin. J. and Auchettl, Katie and Kochanek, Christopher S. and Stanek, K. Z. and Payne, Anna V. and Thompson, Todd A.},
	month = may,
	year = {2020},
	note = {ADS Bibcode: 2020ApJ...894L..10H},
	pages = {L10},
}

@misc{giron_multigroup_2026,
	title = {Multigroup {Radiation} {Diffusion} on a {Moving} {Mesh}: {Implementation} in {RICH} and {Application} to {Tidal} {Disruption} {Events}},
	shorttitle = {Multigroup {Radiation} {Diffusion} on a {Moving} {Mesh}},
	url = {https://ui.adsabs.harvard.edu/abs/2026arXiv260105120G},
	doi = {10.48550/arXiv.2601.05120},
	urldate = {2026-01-15},
	publisher = {arXiv},
	author = {Giron, Itamar and Krief, Menahem and Stone, Nicholas C. and Steinberg, Elad},
	month = jan,
	year = {2026},
	note = {ADS Bibcode: 2026arXiv260105120G},
}

@misc{martire_wind-mediated_2025,
	title = {Wind-mediated {Eddington}-limited emission in a \$10{\textasciicircum}\{4\}{M}\_{\textbackslash}odot\$ {Black} {Hole} {Tidal} {Disruption} {Event}},
	url = {https://ui.adsabs.harvard.edu/abs/2025arXiv251210564M},
	doi = {10.48550/arXiv.2512.10564},
	urldate = {2026-01-15},
	publisher = {arXiv},
	author = {Martire, Paola and Rossi, Elena Maria and Chamberlain Stone, Nicholas and Steinberg, Elad and Kilmetis, Konstantinos and Linial, Itai},
	month = dec,
	year = {2025},
	note = {ADS Bibcode: 2025arXiv251210564M},
}

@article{roth_monte_2015,
	title = {Monte {Carlo} {Radiation}-{Hydrodynamics} {With} {Implicit} {Methods}},
	volume = {217},
	issn = {0067-0049},
	url = {https://ui.adsabs.harvard.edu/abs/2015ApJS..217....9R},
	doi = {10.1088/0067-0049/217/1/9},
	urldate = {2026-01-16},
	journal = {The Astrophysical Journal Supplement Series},
	publisher = {IOP},
	author = {Roth, Nathaniel and Kasen, Daniel},
	month = mar,
	year = {2015},
	note = {ADS Bibcode: 2015ApJS..217....9R},
	pages = {9},
}

@misc{aspegren_emission_2026,
	title = {The {Emission} and {Suppression} of {Line} {Features} in {Luminous} {Transients}},
	url = {https://ui.adsabs.harvard.edu/abs/2026arXiv260100947A},
	doi = {10.48550/arXiv.2601.00947},
	urldate = {2026-01-28},
	publisher = {arXiv},
	author = {Aspegren, Olivia and Kasen, Daniel},
	month = jan,
	year = {2026},
	note = {ADS Bibcode: 2026arXiv260100947A},
}

@incollection{chung_generalized_2016,
	series = {Springer {Series} on {Atomic}, {Optical}, and {Plasma} {Physics}},
	title = {Generalized {Collisional} {Radiative} {Model} {Using} {Screened} {Hydrogenic} {Levels}},
	volume = {90},
	booktitle = {Modern {Methods} in {Collisional}-{Radiative} {Modeling} of {Plasmas}},
	publisher = {Springer, Cham},
	author = {Chung, HK. and Hansen, S.B. and Scott, H.A.},
	year = {2016},
}

@article{wevers_evidence_2019,
	title = {Evidence for rapid disc formation and reprocessing in the {X}-ray bright tidal disruption event candidate {AT} 2018fyk},
	volume = {488},
	issn = {0035-8711},
	url = {https://ui.adsabs.harvard.edu/abs/2019MNRAS.488.4816W},
	doi = {10.1093/mnras/stz1976},
	urldate = {2026-02-19},
	journal = {Monthly Notices of the Royal Astronomical Society},
	publisher = {OUP},
	author = {Wevers, T. and Pasham, D. R. and van Velzen, S. and Leloudas, G. and Schulze, S. and Miller-Jones, J. C. A. and Jonker, P. G. and Gromadzki, M. and Kankare, E. and Hodgkin, S. T. and Wyrzykowski, Ł. and Kostrzewa-Rutkowska, Z. and Moran, S. and Berton, M. and Maguire, K. and Onori, F. and Mattila, S. and Nicholl, M.},
	month = oct,
	year = {2019},
	note = {ADS Bibcode: 2019MNRAS.488.4816W},
	pages = {4816--4830},
}

@article{holoien_unusual_2018,
	title = {The unusual late-time evolution of the tidal disruption event {ASASSN}-15oi},
	volume = {480},
	issn = {0035-8711},
	url = {https://ui.adsabs.harvard.edu/abs/2018MNRAS.480.5689H},
	doi = {10.1093/mnras/sty2273},
	urldate = {2026-02-19},
	journal = {Monthly Notices of the Royal Astronomical Society},
	publisher = {OUP},
	author = {Holoien, T. W.-S. and Brown, J. S. and Auchettl, K. and Kochanek, C. S. and Prieto, J. L. and Shappee, B. J. and Van Saders, J.},
	month = nov,
	year = {2018},
	note = {ADS Bibcode: 2018MNRAS.480.5689H},
	pages = {5689--5703},
}

@article{gezari_x-ray_2017,
	title = {X-{Ray} {Brightening} and {UV} {Fading} of {Tidal} {Disruption} {Event} {ASASSN}-15oi},
	volume = {851},
	issn = {0004-637X},
	url = {https://ui.adsabs.harvard.edu/abs/2017ApJ...851L..47G},
	doi = {10.3847/2041-8213/aaa0c2},
	urldate = {2026-02-19},
	journal = {The Astrophysical Journal},
	publisher = {IOP},
	author = {Gezari, S. and Cenko, S. B. and Arcavi, I.},
	month = dec,
	year = {2017},
	note = {ADS Bibcode: 2017ApJ...851L..47G},
	pages = {L47},
}

@article{reynolds_warm_1995,
	title = {Warm absorbers in active galactic nuclei},
	volume = {273},
	issn = {0035-8711},
	url = {https://ui.adsabs.harvard.edu/abs/1995MNRAS.273.1167R},
	doi = {10.1093/mnras/273.4.1167},
	urldate = {2026-02-19},
	journal = {Monthly Notices of the Royal Astronomical Society},
	publisher = {OUP},
	author = {Reynolds, C. S. and Fabian, A. C.},
	month = apr,
	year = {1995},
	note = {ADS Bibcode: 1995MNRAS.273.1167R},
	pages = {1167--1176},
}

@article{blustin_nature_2005,
	title = {The nature and origin of {Seyfert} warm absorbers},
	volume = {431},
	issn = {0004-6361},
	url = {https://ui.adsabs.harvard.edu/abs/2005A&A...431..111B},
	doi = {10.1051/0004-6361:20041775},
	urldate = {2026-02-19},
	journal = {Astronomy and Astrophysics},
	publisher = {EDP},
	author = {Blustin, A. J. and Page, M. J. and Fuerst, S. V. and Branduardi-Raymont, G. and Ashton, C. E.},
	month = feb,
	year = {2005},
	note = {ADS Bibcode: 2005A\&A...431..111B},
	pages = {111--125},
}

@article{hu_converged_2026,
	title = {Converged {Simulations} of the {Nozzle} {Shock} in {Tidal} {Disruption} {Events}},
	volume = {996},
	issn = {0004-637X},
	url = {https://ui.adsabs.harvard.edu/abs/2026ApJ...996L..21H},
	doi = {10.3847/2041-8213/ae27cc},
	urldate = {2026-02-20},
	journal = {The Astrophysical Journal},
	publisher = {IOP},
	author = {Hu, Fangyi (Fitz) and Mandel, Ilya and Nealon, Rebecca and Price, Daniel J.},
	month = jan,
	year = {2026},
	note = {ADS Bibcode: 2026ApJ...996L..21H},
	pages = {L21},
}

@article{mizumoto_thermally_2019,
	title = {Thermally driven wind as the origin of warm absorbers in {AGN}},
	volume = {489},
	issn = {0035-8711},
	url = {https://ui.adsabs.harvard.edu/abs/2019MNRAS.489.1152M},
	doi = {10.1093/mnras/stz2225},
	urldate = {2026-02-20},
	journal = {Monthly Notices of the Royal Astronomical Society},
	publisher = {OUP},
	author = {Mizumoto, Misaki and Done, Chris and Tomaru, Ryota and Edwards, Isaac},
	month = oct,
	year = {2019},
	note = {ADS Bibcode: 2019MNRAS.489.1152M},
	pages = {1152--1160},
}

@phdthesis{khatami_physics_2024,
	title = {Physics of {Interacting} {Supernova} {Light} {Curves}},
	url = {https://escholarship.org/uc/item/0dt7d765},
	language = {en},
	urldate = {2026-04-28},
	school = {UC Berkeley},
	author = {Khatami, David},
	year = {2024},
}

@article{nugent_synthetic_1997,
	title = {Synthetic {Spectra} of {Hydrodynamic} {Models} of {Type} {Ia} {Supernovae}},
	volume = {485},
	issn = {0004-637X},
	url = {https://ui.adsabs.harvard.edu/abs/1997ApJ...485..812N},
	doi = {10.1086/304459},
	urldate = {2026-04-29},
	journal = {The Astrophysical Journal},
	publisher = {IOP},
	author = {Nugent, Peter and Baron, E. and Branch, David and Fisher, Adam and Hauschildt, Peter H.},
	month = aug,
	year = {1997},
	note = {ADS Bibcode: 1997ApJ...485..812N},
	pages = {812--819},
}

@article{harries_algorithm_2011,
	title = {An algorithm for {Monte} {Carlo} time-dependent radiation transfer},
	volume = {416},
	issn = {0035-8711},
	url = {https://ui.adsabs.harvard.edu/abs/2011MNRAS.416.1500H},
	doi = {10.1111/j.1365-2966.2011.19147.x},
	urldate = {2026-04-30},
	journal = {Monthly Notices of the Royal Astronomical Society},
	publisher = {OUP},
	author = {Harries, Tim J.},
	month = sep,
	year = {2011},
	note = {ADS Bibcode: 2011MNRAS.416.1500H},
	pages = {1500--1508},
}
\bibliographystyle{aasjournal}

\end{CJK*}
\end{document}